\documentclass[12pt]{article}
\usepackage{amsmath}
\usepackage{tensor}
\usepackage{amssymb}
\usepackage{mathrsfs}
\usepackage{tikz}
\usepackage{hyperref}
\usepackage{slashed}
\usepackage{enumerate}
\numberwithin{equation}{section}
\begin{document}
\def\be{\begin{equation}}
\def\bea{\begin{eqnarray}}
\def\ee{\end{equation}}
\def\eea{\end{eqnarray}}
\def\d{\partial}
\def\eps{\varepsilon}
\def\la{\lambda}
\def\b{\bigskip}
\def\nn{\nonumber \\}
\def\p{\partial}
\def\t{\tilde}
\def\h{{1\over 2}}
\def\be{\begin{equation}}
\def\bea{\begin{eqnarray}}
\def\ee{\end{equation}}
\def\eea{\end{eqnarray}}
\def\b{\bigskip}
\def\u{\uparrow}
\newcommand{\comment}[2]{#2}

\begin{center}

\vspace{1cm}
{\Large\bf
Geometries with twisted spheres \\ \ \\ and non--abelian T--dualities
}
\vspace{1.4cm}

{Oleg Lunin\footnote{olunin@albany.edu} and Parita Shah\footnote{pshah3@albany.edu}}\\

\vskip 0.5cm

{\em  Department of Physics,\\
University at Albany (SUNY),\\
Albany, NY 12222, USA
 }

\vskip 0.08cm

\vspace{.2cm}

\end{center}

{\baselineskip 11pt
\begin{abstract}
\noindent
Spectral flow in two--dimensional superconformal field theories is known to correspond to a geometrical mixing between two circles in the gravity dual. We generalize this operation to the geometries which have $SO(k+1)\times SO(k+1)$ isometries with $k>1$ and perform various non--abelian T-dualities of the resulting twisted backgrounds. Combination of non--abelian twists and dualities leads to a new solution generating technique in supergravity, and we apply it to the geometries dual to supersymmetric states in ${\cal N}=4$ super--Yang--Mills theory.
\end{abstract}
}

\newpage
\tableofcontents
\def\be{\begin{equation}}
\def\bea{\begin{eqnarray}}
\def\ee{\end{equation}}
\def\eea{\end{eqnarray}}
\def\d{\partial}
\def\eps{\varepsilon}
\def\la{\lambda}
\def\b{\bigskip}
\def\nn{\nonumber \\}
\def\p{\partial}
\def\t{\tilde}
\def\h{{1\over 2}}
\def\be{\begin{equation}}
\def\bea{\begin{eqnarray}}
\def\ee{\end{equation}}
\def\eea{\end{eqnarray}}
\def\b{\bigskip}
\def\u{\uparrow}
\newpage
\section{Introduction}

Over the last three decades, studies of dualities in string theory have led to many remarkable results, ranging from discovery of D--branes \cite{Polch} and M--theory \cite{Mtheory} to insights into physics of black holes \cite{BHtdual} and AdS/CFT correspondence \cite{AdSCFT}. In particular, combinations of T--dualities and coordinate transformations have played an important role in understanding of charged black holes \cite{SenCvet} and gravity duals of various field theories \cite{TsT,NonCom,TsTdef,BetaDef}. These combinations, known as TsT transformations \cite{TsTdef,TsTold}, involve stringy backgrounds with abelian isometries, and they rely on Buscher rules for the duality maps \cite{OGBuscher}. In recent years significant efforts have been dedicated to extending the solution generating techniques based on duality to the situations when the symmetries of the original manifold are non--abelian \cite{NATD101,NATDAdSCFT}. In this article we will contribute to this effort by exploring networks of non--abelian T--dualities (NATD) for spacetimes containing spheres and by combining NATDs with group rotations to propose non--abelian versions of the TsT transformations. 

Studies of the non--abelian T--duality (NATD) have a long history \cite{NATD80,OldNATD,OldNATD1}, but they have intensified in the last decade as this duality has been applied to produce novel backgrounds relevant for the the AdS/CFT correspondence \cite{NATD101,NATDAdSCFT}. In the abelian case, the procedure for using duality to construct new geometries is rather straightforward: if the original manifold has a $U(1)\times U(1)$ symmetry, one can dualize along the first group, perform a coordinate transformation in the resulting solution by mixing the two cyclic coordinates, and dualize back \cite{TsTold,TsT}. This construction, known as the TsT transformation \cite{TsTdef}, introduces one continuous parameter. In a more general case of $d$ commuting isometries, the dualities and shifts extend to a larger $O(d,d)$ group \cite{SenCvet,RR conventions}, which can be studied naturally in the framework of Double Fields Theory \cite{OddGrp}. In the non--abelian case the situation is more complicated: since NATD generally destroys isometries, the  dualization back to the original frame after a twist presents a significant challenge. However, one can still introduce a counterpart of the TsT transformation in the following way. Starting with a geometry that has $G_1\times G_2$ symmetry, one can either dualize along $G_1$ directly, or do this after mixing $G_1$ and $G_2$. The results of these two procedures would have very low amount of symmetry, but they would be related by a non--abelian version of the TsT transformation. The goal of the present article is to introduce this solution generating technique and to explore its consequences. After presenting the general constructions, we will focus on examples coming from products of two-- and 
three--spheres, i.e., on dualities along twisted $SO(3)\times SO(3)$ and $SO(4)\times SO(4)$ isometries.

There are two distinct procedures for performing NATD: one is applicable to group manifolds \cite{NTDgrp,1012}, and the other is used to dualize along cosets \cite{1104}.  Both methods have been used in the past to generate new solutions starting with backgrounds like $AdS_3\times S^3\times T^4$ and $AdS_5\times S^5$ \cite{NATDAdSCFT,NATD101} and to interpret the results as interesting new solutions in M-theory \cite{TDUplift}. In some cases, a given background can be dualized by either a group or a coset prescription, and the results of these two procedures are rather different. One of the goals of the present article is to explore the interplay between the two distinct duality procedures when both of them are available. The simplest example of the relevant space is a three--dimensional sphere $S^3$, which can be viewed either as the $SU(2)$ group manifold or as the $SO(4)/SO(3)$ coset. The corresponding dualities and relations between them will be analyzed in section \ref{SecS3}. 

While a single three--dimensional sphere constitutes an excellent model for exploring interplay between non--abelian T--dualities, the situation becomes even more interesting when the starting geometry contains a product of spheres, e.g., $S^3\times S^3$. In this case one can introduce 
a non--abelian generalization of the TsT transformations by mixing the spheres before performing dualities. We will explore this procedure in sections \ref{SecSpecFlow}-\ref{SecSfTdCoset}. The largest class of explicit supergravity solutions containing the 
$S^3\times S^3$ product is known as bubbling geometries \cite{LLM}, and article \cite{ILCS12} explored the action of NATDs on one of the spheres present in these metrics. The bubbling solutions describe gravity duals of supersymmetric states in the 
${\cal N}=4$ super--Yang--Mills theory, and they have lower--dimensional counterparts, so called fuzzball geometries \cite{fuzzD1D5,fuzz}, which correspond to supersymmetric states in two--dimensional CFT \cite{D1D5CFT}. In the latter case, field theory admits a very interesting operation of spectral flow \cite{SchwSeib,SpecFlow} which changes boundary conditions for fermions and allows one to go between the NS and the Ramond sectors. On the gravity side, this spectral flow corresponds to a shift in angular coordinates which mixes two $U(1)$ symmetries present in the geometries\footnote{Spectral flow can also be defined for more general fuzzballs which don't have $U(1)\times U(1)$ symmetries, but discussion of this operation is beyond the scope of our article.}. Given importance of the spectral flow described by twisting $U(1)\times U(1)$ symmetries of the fuzzball solutions, it is natural to ask whether a similar operation can be introduced for bubbling geometries. We define the relevant procedure in section \ref{SecSpecFlow} by performing a twist between two blocks of the $SO(4)\times SO(4)$ symmetry. This twist is also generalized to geometries with $SO(k+1)\times SO(k+1)$ for arbitrary $k$. In section \ref{SecSfTdGroup} we combine the twisting with NATDs to introduce a new solution generating technique and apply it to the full family of bubbling solutions.

\bigskip

This paper has the following organization. In section \ref{SecReview} we review the procedures associated with NATD for groups and cosets to use them in the remaining part of the article. In section \ref{SecS3} we focus on geometries which contain an $S^3$ factor, carry out various non--abelian dualities along this sphere, and find interesting relations between NATDs performed over groups and cosets. In section \ref{SecSpecFlow} we introduce a counterpart of the spectral flow operation for geometries which have products of spheres and focus on specific examples of metrics with two $S^3$ factors. While the resulting twisted geometries are equivalent to their untwisted counterparts, at least in the bosonic sector, they can be used to produce new nontrivial solutions by application of NATD. This new solution generating technique is explored in sections \ref{SecSfTdGroup} and \ref{SecSfTdCoset}. Specifically, in section \ref{SecSfTdGroup} we apply the combination of twist and NATD to generate four new families of supergravity solutions, and each of these families is parameterized by one harmonic function of three variables. Section \ref{SecSfTdCoset} outlines the procedure of applying twist--NATD procedure to spaces containing $S^k\times S^k$ for arbitrary k and presents explicit solutions for the $k=2$ case. Some technical details of calculations are presented in appendices.

\section{Review of non-abelian T-duality}
\label{SecReview}
In this section we review the procedure for performing non-abelian T-dualities (NATD) on a non-linear $\sigma$-model \cite{OldNATD1}. Specifically, we summarize the prescriptions for NATD along a group manifold \cite{1012} and along a coset \cite{1104}. The resulting formulas will be used in the remaining parts of this article.

We begin with considering a non-linear $\sigma$--model on a manifold describing a Lie group $g$. To describe the groups and cosets in a uniform way, we will make only the symmetry under the left action of $G$ explicit, and the most general action with this property can be written as 
\begin{equation}\label{basic sigma model with E}
S=\int d^2\sigma \left[E_{ij}L^i_+L^j_-+Q_{\mu\nu}\d_+Y^\mu \d_-Y^\nu+Q_{\mu i}\d_+Y^\mu L_-^i+
Q_{i\mu }L_+^i \d_-Y^\mu\right]\,.
\end{equation}
Here and below
\bea
L^i_\mu=-i\operatorname{Tr}(t^ig^{-1}\partial_\mu g)\quad \mbox{and}\quad L^i_\pm=L^i_\mu \partial_\pm X^\mu
\eea
denote the left invariant Maurer-Cartan forms in the target space and on the worldsheet. Our conventions for generators are summarized in Appendix \ref{AppAsub1}. By construction, action (\ref{basic sigma model with E}) is invariant under the left action $g\rightarrow h_L g$ of the group for any matrix $E_{ij}$, but for $E_{ij}=E\delta_{ij}$ and $Q_{\mu i}=Q_{i\mu}=0$ the symmetry is enhanced to 
$G_L\times G_R$. To dualize the action (\ref{basic sigma model with E}) along $G_L$, one  gauges this isometry and introduces Lagrange multipliers to ensure that the gauge potential has vanishing field strength\footnote{To avoid unnecessary complications, in this intermediate expression we set 
$Q_{\mu i}=Q_{i\mu}=0$. The normalizations of the gauge field and its strength follow \cite{NATD101} $D_{\pm}g=\partial_\pm g-A_\pm$, $F_{+-}=-[D_+,D_-]$.}:
\bea\label{ActPreDual}
S_{\text{}}=\int d^2\sigma \left[-E_{ij}\operatorname{Tr} (t^ig^{-1}D_+g)\operatorname{Tr}(t^jg^{-1}D_-g)    -i\operatorname{Tr}(vF_{+-})+Q_{\mu\nu}\d_+Y^\mu \d_-Y^\nu\right]\,.
\eea
The original action (\ref{basic sigma model with E}) can be recovered by integrating out Lagrange multipliers, while the dual theory is obtained by  integrating out the gauge fields. Specifically, observing that the action (\ref{ActPreDual}) has a gauge symmetry parameterized by $h_L(\sigma^+,\sigma^-)\in G$:
\bea
g\rightarrow h_L^{-1}g, \quad v\rightarrow h_L^{-1}vh_L, \quad A_\pm\rightarrow  h_L^{-1}A_\pm h_L-h_L^{-1}\partial_\pm h_L  
\eea
one can fix this symmetry by setting $g=\mathbb{I}$. Then integrating out the gauge fields, one gets the action in terms of the dual coordinates $v$ \cite{NATD101}:
\bea\label{basic dual}
S_{\text{dual }}=\int d^2\sigma \left[(\partial_+v_i+\d_+ Y^\mu Q_{\mu i})(M^{-1})^{ij}
(\partial_-v_j-Q_{j\nu}\d_-Y^\nu)+Q_{\mu\nu}\d_+Y^\mu \d_-Y^\nu\right]\,.
\eea
Matrices $M_{ij}$ and $f_{ij}$ are defined in terms of $E_{ij}$ and structure constants $f_{ij}{}^k$  of the group $G$:
\begin{equation}\label{dualMij}
M_{ij}=(E_{ij}+f_{ij}), \quad f_{ij}\equiv f_{ij}{}^kv_k\,.
\end{equation}
The metric and the Kalb--Ramond field of the dual theory are encoded in the matrix $M$, and the new dilaton is given by
\begin{equation}\label{basic dilaton}
\phi_{\text{dual}}=\phi-\frac{1}{2} \ln({\operatorname{det}M})\,.
\end{equation}
Expressions (\ref{basic dual}) and (\ref{basic dilaton}) give all NS--NS fields of the dual geometry.

In order to determine the Ramond--Ramond (RR) forms of the dual, one begins with defining frames for the metric encoded by (\ref{basic dual}):
\bea\label{DualFrames}
&&{\hat e}^a_+=-{\kappa^a}_j(M^{-1})^{ij}(dv_i+d Y^\mu Q_{\mu i})+{\la^a}_\mu dY^\mu,\nn
&&{\hat e}^a_-={\kappa^a}_i(M^{-1})^{ij}(dv_j- Q_{j\mu}d Y^\mu)+{\la^a}_\mu dY^\mu\,.
\eea
Here matrices $\kappa$ and $\lambda$ are defined by the frames for the original action 
(\ref{basic sigma model with E}):
\bea
e^a=\tensor{\kappa}{^a_i}L^i+\tensor{\lambda}{^a_\mu}dY^\mu,\quad e^A=\tensor{e}{^A_\mu}dY^\mu\,,
\eea
The frames ${\hat e}^a_-$ and ${\hat e}^a_+$ are related by a local Lorentz transformation with matrix $\Lambda$ given by
\begin{equation}\label{lambda group}
\Lambda_i{}^j=-(\kappa^{-T}MM^{-T}\kappa^T)_i{}^j \,. 
\end{equation}
Action of this transformation on spinors is given by matrix $\Omega$ which is defined by \cite{SfetRR,RR conventions}
\begin{equation}\label{Omdefn}
\Omega^{-1}\Gamma^{i}\Omega=\Lambda^i{}_j\Gamma^j  \,.  
\end{equation}
Note that the Lorentz transformation changes parity when dimension of group $G$ is odd and it does not when the dimension is even. Therefore for groups with odd dimension the duality maps type-IIA and type-IIB theories into each other. The RR forms are nicely summarized in terms of a bi-spinor $P$ defined by
\begin{equation}\label{RRproj}
\text{IIA: }\hat{P}=\frac{e^{\hat{\phi}}}{2}\sum^{5}_{n=0}   \frac{1}{(2n)!}\slashed{\hat{F}}_{2n},\quad \text{IIB: }P=\frac{e^{{\phi}}}{2}\sum^{4}_{n=0}   \frac{1}{(2n+1)!}\slashed{{F}}_{2n+1}\,,
\end{equation}
where $\slashed{F}_m=\Gamma^{{\mu_1}{\mu_2}\ldots{\mu_m}}F_{{\mu_1}{\mu_2}\ldots{\mu_m}}$. For odd $\operatorname{dim}(G)$, the dualilty map for the RR forms is given by \cite{OldNATD1,SfetRR}
\begin{equation}\label{RR trans defn}
\hat{P}=P\Omega^{-1}\,,    
\end{equation}
where the matrix $\Omega$ is defined by $(\ref{Omdefn})$. 

\bigskip

An extension of the duality procedure to coset spaces was introduced in \cite{1104}, and it can be summarized as follows. Coset models are described by the action similar to (\ref{basic sigma model with E})\footnote{To avoid unnecessary clutter, we focus only on the $E$--part of the action and drop terms with $Q$.},
\begin{equation}\label{coset action}
S=\int d^2\sigma (E_0)_{\alpha\beta}L^\alpha_\mu L^\beta\partial_+X^\mu\partial_-X^\nu
\end{equation}
with a summation index $\alpha$ running over the $G/H$ coset rather than the entire group 
$G$. Such models can be recovered from  (\ref{basic sigma model with E}) by starting with an $E$--matrix of the form $E=\operatorname{diag}(E_0,\lambda \mathbb{I}_{\operatorname{dim}(H)})$ and taking the limit $\lambda\rightarrow 0$.  Then the procedure (\ref{basic dual}) can still be used to obtain the dual, but $\operatorname{dim}(H)$ of the Lagrange multipliers $v_i$ must be fixed by a rotation from a subgroup $H$. The dilaton is still given by (\ref{basic dilaton}). 

To obtain the RR forms for the coset case, the matrix $\Lambda$ of the Lorentz transformation  is constructed by defining a $\operatorname{dim}(G/H)$ square matrix $N$ via taking the $\lambda\rightarrow 0$ limit in various components of (\ref{basic dual}) and fixing the gauge for the subgroup:
\begin{equation}\label{N definition}
\begin{aligned}
L^\alpha_+&=(M^{-1})^{i\alpha}\partial_+v_i=   N^{\alpha\beta}_+\partial_+x_\beta\,, \\
L^\alpha_-&=-(M^{-1})^{\alpha i}\partial_-v_i=   N^{\alpha\beta}_-\partial_-x_\beta\,. \\
\end{aligned}
\end{equation}
The matrix describing the Lorentz transformation is given by
\begin{equation}\label{LaCoset}
\Lambda_\alpha{}^\beta=(N_+N_-^{-1})_\alpha{}^\beta \,.   
\end{equation} 
and it defines $\Omega$ through the relation (\ref{Omdefn}). 
The RR forms are constructed using the transformation rule (\ref{RR trans defn}). It should be noted that here although the duality is performed along the coset $G/H$, the parity of the dual theory is governed by $\operatorname{dim}(G)$ rather than the $\operatorname{dim}(G/H)$. Some recent applications of NATD can be found in \cite{NATDapp}.

\section{Web of non--abelian dualities along $S^3$}
\label{SecS3}

As reviewed in the previous section, dualization along groups and cosets gives rise to different NATD prescriptions. This raises an interesting question about relation between these two procedures when both groups and cosets are present. In this section we will explore such relations for geometries containing a three--dimensional sphere $S^3$, which can be viewed either as an $SU(2)$ group manifold or as an $SO(4)/SO(3)$ coset. We will demonstrate that dualization along $SU(2)$ produces a manifold with residual symmetries, and an additional T--duality along these directions results in a geometry equivalent to the non--abelian T dual of $S^3$ viewed as the $SO(4)/SO(3)$ coset. Implications of this relation will be discussed in the end of this section. 

Let us consider solutions of type IIB supergravity that contain an $S^3$ factor and which are supported by the five--form flux:
\bea\label{S3start}
ds^2=ds_7^2+A\, d\Omega_3^2,\quad F_5=G_2\wedge d^3\Omega+dual\,.
\eea
This large class of geometries contains the flat ten--dimensional space and $AdS_5\times X$, which can be written as
\bea
ds^2=L^2\left[-(1+\rho^2)dt^2+\frac{d\rho^2}{1+\rho^2}+\rho^2 d\Omega_3^2\right]+ds_X^2,\quad F_5=\rho^3 dt\wedge d\rho\wedge d^3\Omega+dual\,.
\eea
Additional examples include so--called bubbling geometries with $SO(4)\times SO(4)$ isometries which will be studied in detail in the next section. Here we will not make any assumptions about the seven dimensional manifold $X$ in (\ref{S3start}) and perform non--abelian T--dualities along the three--dimensional sphere and its ingredients. While any $d$--dimensional sphere can be viewed as a coset $SO(d+1)/SO(d)$, $S^3$ can also be interpreted as an $SU(2)$ group manifold using the relation
\bea\label{OmegaAsSU2}
d\Omega_3^2=-\frac{1}{4}\mbox{tr}(g^{-1}dgg^{-1}dg)\,,
\eea
where $g$ is a unitary $2\times 2$ matrix parameterized by three coordinates. This peculiar feature of $S^3$ gives rise to three distinct dualization procedures:
\begin{enumerate}[(a)]
\item Dualization along the $SO(4)/SO(3)$ coset. This procedure will be performed in section \ref{SecSubCoset}, and it results in a geometry without isometries, apart from those inherited 
from $ds_7^2$.
\item Dualization along the group $SU(2)$. This procedure will be performed in section \ref{SecSubSU2}, and it results in a geometry with a residual $SU(2)$ isometry. Specifically, the metric (\ref{OmegaAsSU2}) is invariant under $SU(2)_L\times SU(2)_R$ which acts as 
\bea\label{ActSU2LR}
g\rightarrow h_L g h_R.
\eea
Dualization along $h_L$ preserves the symmetry under $h_R$.
\item The result of the option (b) can be further dualized along the $SO(3)/SO(2)$ to produce a geometry without isometries, apart from those inherited from $ds_7^2$. In section \ref{SecSubSU2dbl} we will demonstrate that this final answer is related to the result of option (a) by an interesting change of variables.
\end{enumerate}
The physical interpretation of the map appearing in option (c) will be discussed in sections \ref{SecSubSU2dbl} and \ref{SecSubAbel}.

\subsection{Duality along the $SO(4)/SO(3)$ coset}
\label{SecSubCoset}

We begin with viewing the sphere in (\ref{S3start}) as an $SO(4)/SO(3)$ coset and dualizing along it using the prescription (\ref{coset action}). To do so, we gauge the $G=SO(4)$ symmetry and fix the residual degrees of freedom describing $H=SO(3)$. Note that although the dimension of the sphere is odd, the resulting T dual has the same parity as the original model (\ref{S3start}) since $G=SO(4)$ has even dimension.

Writing (\ref{S3start}) in terms of the coset element $g$ and comparing the result with (\ref{coset action}), we conclude that 
$E_0=A\,\mathbb{I}$. Therefore, in the limit of vanishing $\lambda$, the matrix $M$ describing the dual geometry (\ref{basic dual}) for $SO(4)$ is given by \cite{1104}
\begin{equation}
M=\left[
\begin{array}{cccccc}
A&-v_4&-v_5&v_2&v_3&0\\
v_4&A&-v_6&-v_1&0&v_3\\
v_5&v_6&A&0&-v_1&-v_2\\
-v_2&v_1&0&0&-v_6&v_5\\
-v_3&0&v_1&v_6&0&-v_4\\
0&-v_3&v_2&-v_5&v_4&0\\
\end{array}
\right]\,,    \nonumber
\end{equation}
where $v_i$ are the Lagrange multipliers. To describe the dual of the coset, we need to gauge fix the subgroup $H=SO(3)$ which acts by a simultaneous rotation in two three--dimensional spaces spanned by $(v_1,v_2,v_3)$ and $(v_4,v_5,v_6)$. Fixing the gauge by setting $v_1=v_2=v_6=0$ and treating $(v_3,v_4,v_5)=(x_1,x_2,x_3)$ as coordinates on the dual manifold, we arrive at the final matrix $M$:
\begin{equation}\label{matrix M for coset ads}
M=\left[
\begin{array}{cccccc}
A&-x_2&-x_3&0&x_1&0\\
x_2&A&0&0&0&x_1\\
x_3&0&A&0&0&0\\
0&0&0&0&0&x_3\\
-x_1&0&0&0&0&-x_2\\
0&-x_1&0&-x_3&x_2&0\\
\end{array}
\right]  
\end{equation}
This result leads to the frames (\ref{DualFrames}): 
\bea\label{frames}
e^1&=&\frac{A x_2}{\sqrt{A}x_1x_3}dx_2+\frac{Ax_3}{\sqrt{A}x_1x_3}dx_3,\quad 
 e^2=\frac{x_2^2-x_1^2}{\sqrt{A}x_1x_3}dx_2+\frac{x_2x_3}{\sqrt{A}x_1x_3}dx_3,\nn
 e^3&=&\frac{1}{\sqrt{A}}dx_1+\frac{x_2x_3}{\sqrt{A}x_1x_3}dx_2+\frac{x_3^2}{\sqrt{A}x_1x_3}dx_3,
\eea
and to the metric 
\bea\label{MetrCosetDual}
 d{s}^2_{\text{coset}}=ds_7^2+\frac{dR^2}{Ax_1^2}
 +\bigg[dx_3\bigg(\frac{x_2}{x_1}\bigg)+dx_2\frac{(x_2^2-x_1^2)}{x_1x_3}\bigg]^2\frac{1}{A}+\bigg[\frac{dx_3}{x_1}+dx_2\bigg(\frac{x_2}{x_1x_3}\bigg)\bigg]^2A\,.
\eea
To make the last expression shorter, we defined a convenient function $R$:
\bea
R^2=(x_1)^2+(x_2)^2+(x_3)^2.
\eea
Duality along the coset does not generate a B field, but it produces a nontrivial dilaton: 
\bea\label{coset dilaton}
e^{2\phi_{\text{coset}}}=\frac{1}{(Ax_1x_3)^2},\quad B_{\text{coset}}=0.
\eea
Equations (\ref{MetrCosetDual}) and (\ref{coset dilaton}) summarize all the NS--NS fields for the dual background.

\bigskip 

To construct the Ramond--Ramond fields, we begin with extracting matrices $N_\pm$ defined 
in (\ref{N definition}) which lead to the Lorentz rotation $\Lambda$ and matrix $\Omega$ given by (\ref{LaCoset}) and (\ref{Omdefn}). The results read
\bea
N_\pm =\frac{1}{x_1x_3 A}\begin{bmatrix}
0&x_2&x_3\\
0&\pm(x_2^2-x_1^2)&\pm x_2x_3\\
\pm x_1x_3&\pm x_2x_3&\pm x_3^2
\end{bmatrix}, \ \Lambda=\operatorname{diag}(1,-1,-1),\ \Omega=-\Gamma_2\Gamma_3\,.
\eea
Here indices $1,2,3$ represent the directions of $S^3$ and its dual. To proceed we introduce a convenient notation for the gamma matrices on the seven--dimensional space $ds_7^2$ by splitting $\Gamma_{11}$ into a product of $\Gamma_G$ pointing along the direction of the flux and $\Gamma_\perp$ pointing in the orthogonal directions:
\bea
\slashed{{G}}_2=g_2\Gamma_G,\quad \Gamma^{0456789}=\Gamma_G\Gamma_\perp\,.
\eea 
Then the dual RR forms are obtained using (\ref{RRproj}) and (\ref{RR trans defn}) with $\hat{\phi}=\phi_{\text{coset}}$ and 
\bea
P=\Gamma_\perp-\Gamma^{123}\Gamma_G\,.
\eea
The result reads 
\bea\label{cosetRR}
{\hat F}^{(3)}=2Ax_1x_3~e^1\wedge \frac{G_2}{A^{3/2}}
={2}(x_2dx_2+{x_3}dx_3)\wedge G_2,\quad \hat{F}_7=-(\star \hat{F}_3).
\eea
Equations (\ref{MetrCosetDual}), (\ref{coset dilaton}), (\ref{cosetRR}) constitute the final results of this subsection. Note that the resulting geometry has no isometries, as one would expect for duals of coset spaces. In the next two subsections we will explore alternative dualizations along $S^3$.

\subsection{Duality along the $SU(2)_L$}
\label{SecSubSU2}
Let us now view the background (\ref{S3start}) as a product of $X_7$ and a group manifold (\ref{OmegaAsSU2}) and dualize along this $SU(2)$ using the procedure (\ref{basic dual}) and (\ref{RR trans defn}).  Since the group is odd--dimensional, the dual model is described by the  type-IIA theory.

\bigskip 

Using the general expressions (\ref{basic dual}) to construct the matrix $M$ for the dual background ,
\begin{equation}
 M=\begin{bmatrix}
A&x_3&-x_2\\
-x_3&A&x_1\\
x_2&-x_1&A
\end{bmatrix}\,,
\end{equation}
and extracting the NS--NS fields from it, we find 
\begin{equation}\label{Group1}
\begin{aligned}
 ds^2_{\text{dual 1}}&=ds_7^2 +\frac{1}{4}\left[\frac{dr^2}{A}+\frac{r^2A}{A^2+r^2}d\Omega_2^2\right]\,,\\
B_{\text{dual 1}}&=\frac{1}{4}\frac{\epsilon r^3}{A^2+r^2}\operatorname{Vol}(S^2),\quad\phi_{\text{dual 1}}=-\frac{1}{2}\ln[\frac{A(A^2+r^2)}{64}]\,.\\
\end{aligned}   
\end{equation}
Here $(r,\Omega_2)$ are the spherical coordinates in three--dimensional space spanned by $(x_1,x_2,x_3)$, and the parameter $\epsilon=\pm 1$ indicates the choice between dializing along one of the two $SU(2)$ present in the parent theory\footnote{Recall that $SO(4)\sim SU(2)\times SU(2)$.}. The remaining $SU(2)$ is still intact, and it is encoded in the $S^2$ factor in (\ref{Group1}). 

To determine the RR fields supporting the background (\ref{Group1}), we begin with finding the matrix of the Lorentz transformation defined by (\ref{lambda group}):
\begin{equation}\label{lambda su2}
\tensor{\Lambda}{_i^j}=\frac{-A^2+r^2}{A^2+r^2}\delta_{ij}-\frac{2(x_ix_j+A\epsilon_{ijk}x_k)}{A^2+r^2}\,.
\end{equation}
Further, using the definition (\ref{Omdefn}) of $\Omega_{\text{dual 1}}$, we find
\begin{equation}\label{Om su2}
\Omega_{\text{dual 1}}=\Gamma_{11}\Tilde{\Omega},\quad \Tilde{\Omega}=\frac{A\Gamma_{123}+\bold{(x\cdot\Gamma)}}{\sqrt{A^2+r^2}}\,.   
\end{equation}
Appearance of matrix $\Gamma_{11}$ is a general feature of transitioning between type IIB and IIA theories by dualization along an odd--dimensional manifold. 
Application of the general prescription (\ref{RR trans defn}) gives
\begin{equation}
\begin{aligned}
\frac{e^{{\phi}_{\text{dual 1}}}}{2}\bigg(\frac{1}{2}\slashed F_2+\frac{1}{4!}\slashed F_4+\frac{1}{6!}\slashed F_6+\frac{1}{8!}\slashed F_8\bigg)=\bigg(\Gamma_\perp-\Gamma^{123}\Gamma_G\bigg)\Omega^{-1}_{\text{dual 1}}\,,\\
\end{aligned}    
\end{equation}
and the final expressions are 
\bea\label{Group1RR}
F_4=-\frac{8r^3}{A^2+r^2}G_2\wedge \operatorname{Vol}(S^2),\quad  F_2=-8 G_2,\quad F_6=\star F_4,\quad F_8=-\star F_2\,.
\eea
Note that since the original background does not have 3-forms or 1-forms, there is no 0--form in the dual theory. To obtain (\ref{Group1RR}) we used the explicit expressions for the frames $(e^2,e^2,e^3)$, as well as some algebraic relations between them \cite{1012}:
\bea
e^i=\frac{1}{2\sqrt{A(A^2+r^2)}}\left[Adx^i+\frac{x^i}{r}(\sqrt{A^2+r^2}-A)dr\right],\ 
x_ie^i=\frac{1}{2\sqrt{A}}x_idx^i=\frac{1}{2\sqrt{A}}rdr\,.\nonumber
\eea
Expressions (\ref{Group1}), (\ref{Group1RR}) constitute the results of this subsection. Since the final background has an $SO(3)$ isometry, it is interesting to perform an additional duality along it and compare the outcome with the results of the previous subsection. This will be our next step.

\subsection{Duality along the remaining $S^2$}
\label{SecSubRemS2}
\label{SecSubSU2dbl}
In this subsection we will perform a T-duality along the two--dimensional sphere in (\ref{Group1}). In contrast to the $S^3$, which can be viewed either as a group manifold or as a coset, now there is no choice but to view $S^2$ as $SO(3)/SO(2)$ and to perform the appropriate NATD. Although the dimension of this coset is even, the T--duality rules imply that the chirality of fermions flips since $SO(3)$ is three--dimensional, so the dual of the geometry (\ref{Group1}) is governed by the type IIB theory. Therefore, it would be instructive to compare the result with expressions (\ref{MetrCosetDual}), (\ref{coset dilaton}), and (\ref{cosetRR}).

To perform the duality, we begin with reading off the initial $E$ matrix by comparing 
(\ref{Group1}) with (\ref{coset action}):
\begin{equation}
E_{ij}=(\mathbb{E}_{\alpha\beta},\lambda),\quad \text{where}\quad \mathbb{E}
_{\alpha\beta}=\frac{r^2A}{A^2+r^2}\delta_{\alpha\beta}+{(B_{\text{dual 1}})}_{\alpha\beta} \,.  
\end{equation}
This leads to the matrix $M$, which is given it terms of $E$ and three Lagrange multipliers $v_i$ by the general relation (\ref{basic dual}):
\begin{equation}\label{MmatrSU2p}
\begin{aligned}
M_{ij}&=\begin{bmatrix}
\frac{Ar^2}{A^2+r^2}&\frac{r^3}{A^2+r^2}+v_3&-v_2\\
-\frac{r^3}{A^2+r^2}-v_3&\frac{Ar^2}{A^2+r^2}&v_1\\
v_2&-v_1&0
\end{bmatrix}
\end{aligned}   \,. 
\end{equation}
The $SO(2)$ subgroup acts by rotating $(v_1,v_2)$ and leaving $v_3$ invariant, so an element of the $SO(3)/SO(2)$ coset can be specified by imposing the gauge $v_1=0$ and setting $(v_2,v_3)\equiv (y_1,y_2)$ in (\ref{MmatrSU2p}). Then equation (\ref{DualFrames}) leads to the frames 
\bea
e^1=-\frac{dy_2}{2y_1}\sqrt{\frac{A r^2}{A^2+r^2}},\quad
e^2=\frac{1}{2}\left[\sqrt{\frac{A^2+r^2}{A r^2}}dy_1+dy_2\frac{(r^3+(A^2+r^2)y_2)}{\sqrt{A r^2(A^2+r^2)}y_1}\right]\,,
\eea
and to the NS--NS fields for the dual background 
\bea\label{DualSU2p}
d{s}^2_{\text{dual 2}}&=&ds_7^2+\frac{dr^2}{4A}
+\frac{Ar^2}{4(A^2+r^2)}\left\{\frac{dy_2^2}{y_1^2}+\bigg[dy_1\frac{(A^2+r^2)}{Ar^2}+dy_2\frac{r^3+(A^2+r^2)y_2}{Ar^2y_1}\bigg]^2\right\}\,,\nn
e^{\phi_{\text{dual 2}}}&=&\frac{8}{Ary_1},\quad B=0\,.
\eea
To construct the Ramond--Ramond fields, we extract the $N_\pm$ matrices defined by (\ref{N definition}) and the corresponding matrices $\Lambda$, $\Omega_{\text{dual 2}}$:
\bea
 N_\pm=\frac{1}{Ar^2y_1}\begin{bmatrix}
0&-Ar^2\\
\pm y_1(A^2+r^2)&\pm[r^3+(A^2+r^2)y_2]
\end{bmatrix},\ \Lambda=\operatorname{diag}(1,-1),\
\Omega_{\text{dual 2}}=\Gamma_{11}\Gamma_{2}.\nonumber
\eea
Application of the duality rules (\ref{RRproj})--(\ref{RR trans defn}) leads to the RR forms for the dual background 
\bea\label{DualSU2pRR}
 F^{(3)}=-8(y_1dy_1+y_2dy_2)\wedge G_2,\quad 
F^{(7)}=-\star F^{(3)}\,.
\eea
We have a type-IIB theory with a vanishing B-field and no remaining isometries. 

\bigskip

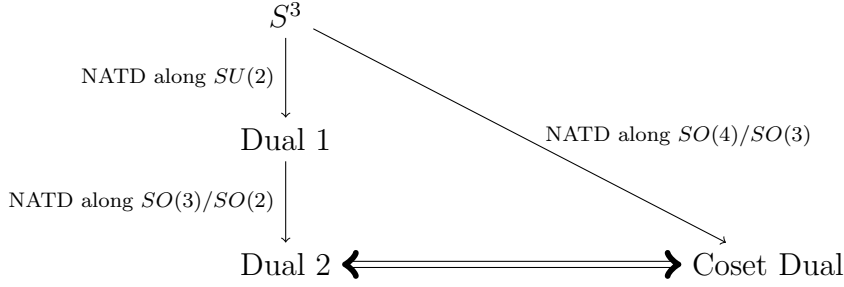
\begin{figure}
\[
\begin{tikzpicture}[scale=4.75]
\node (A) at (0,0.7) {$S^3$};
\node (F) at (0,0) {Dual 2};
\node (C) at (0,0.35) {Dual 1};
\node (D) at (1.35,0) {Coset Dual};
\draw[->] (A) -- (C) node[midway,left]{\scriptsize{NATD along $SU(2)$}};
\draw[->] (C) -- (F) node[midway,left]{\scriptsize{NATD along $SO(3)/SO(2)$}};
\draw[->] (A) -- (D) node[midway,right]{\scriptsize{~~NATD along $SO(4)/SO(3)$}};
\draw[<->,double equal sign distance] (F) -- (D) node[above]{};
\end{tikzpicture}\]
\caption{Network of dualities acting on the $S^3$ factor in the geometries (\ref{S3start}).}
\label{FigTriangle}
\end{figure}

Let us compare the geometry (\ref{DualSU2p})--(\ref{DualSU2pRR}), obtained by dualizing the three--sphere  along the $SU(2)$ group manifold followed by duality along $SO(3)/SO(2)$, with the background (\ref{MetrCosetDual}) obtained by a direct dualization along $S^3$ viewed as the $SO(4)/SO(3)$ coset. Although the geometries (\ref{MetrCosetDual}) and (\ref{DualSU2p})  look rather different, it turns out that there is a simple relation between them. Our final result is summarized in figure \ref{FigTriangle}.

We begin with comparing the NS--NS fields for the backgrounds (\ref{MetrCosetDual})  and (\ref{DualSU2p}). Direct calculation indicates that these geometries are related by a coordinate transformation
\bea\label{SU2map}
x_1=\frac{1}{4}\sqrt{y_1^2+(2r+y_2)^2},\quad x_2=\frac{y_1^2+y_2(2r+y_2)}{4\sqrt{y_1^2+(2r+y_2)^2}},\quad x_3=\frac{r y_1}{2\sqrt{y_1^2+(2r+y_2)^2}}\,.
\eea
The RR fields, (\ref{cosetRR}) and (\ref{DualSU2pRR}), are matched under this map as well. 

The triangle diagram shown in figure \ref{FigTriangle} is a consequence of a peculiar property of the three--dimensional sphere which can be viewed either as a group manifold or as a coset. It is interesting to note that the success in matching (\ref{MetrCosetDual})  and (\ref{DualSU2p}) stems from a somewhat non--intuitive property of a coset duality: even though the coset $SO(4)/SO(3)$ is three-dimensional, the duality maps type IIB theory into type IIB since the group has dimension six. Therefore, duality along the coset effectively correspond to six T--dualities, which is the same number of transformations that one obtains by subsequent dualization along $SU(2)$ and $SO(3)/SO(2)$. To make this counting more intuitive, in the next subsection we will discuss the abelian limit of the dualization procedure and the triangle diagram \ref{FigTriangle}.

\subsection{Abelian limit: T--dualities on a torus}
\label{SecSubAbel}

In sections \ref{SecSubCoset}--\ref{SecSubSU2dbl} we studied various T--dualities along the three--dimensional sphere and relations between them, and it is instructive to look at the limit of infinite radius of the sphere. In this case, NATD reduces to its Abelian version, and the standard Buscher's rules are recovered. 

Let us go back to the geometry (\ref{S3start}) and consider a limit of large warp factor $A$. To do that, we write
\bea\label{AflatLim}
A=\Lambda^2 a,\quad G_2=\Lambda^3 g_2
\eea
and send $\Lambda$ to infinity assuming that $a$ and $g_2$ remain fixed. This limit gives
\bea
&&A\, d\Omega_3^2=A(d\theta^2+\sin^2\theta d\Omega_2^2)\rightarrow
a(d{\tilde\theta}^2+{\tilde\theta}^2 d\Omega_2^2)=a\,d{\bf x}\cdot d{\bf x},\quad {\tilde\theta}=\Lambda\theta,\nn
&&G_2\wedge d^3\Omega\rightarrow g_2\wedge {\tilde\theta}^2 d{\tilde\theta}\wedge d^2\Omega=g_2\wedge d^3 x\,,
\eea
leading to a counterpart of (\ref{S3start})
\bea\label{S3startFlat}
ds^2=ds_7^2+a\, dx_{3d}^2,\quad F_5=g_2\wedge d^3x+dual\,.
\eea
Application of the same limit to the metric  (\ref{Group1}) obtained after dualization along $SU(2)$ gives
\begin{equation}\label{Group1lim}
\begin{aligned}
 ds^2_{\text{dual 1}}=ds_7^2 +\frac{1}{a}\left[d{\hat r}^2+{\hat r}^2d\Omega_2^2\right],
 \quad
B_{\text{dual 1}}=0,\quad e^{2\phi_{\text{dual 1}}}=\frac{1}{a^3}\,.\\
\end{aligned}   
\end{equation}
As expected, this geometry describes dualization of (\ref{S3startFlat}) along the three--dimensional torus. Now, let us consider the result obtained after the second step of the 2-step duality i.e. dualizing the $S^2$ with respect to a coset and apply the flat space limit for that result (\ref{DualSU2p})  as well: 
\bea\label{DualSU2pFlat}
d{s}^2_{\text{dual 2}}=ds_7^2+\frac{d{\hat r}^2}{4a}
+\frac{{\hat r}^2}{a}\frac{dy_2^2}{4y_1^2}+\frac{a}{4{\hat r}^2}\bigg[dy_1+\frac{y_2dy_2}{y_1}\bigg]^2,\
e^{2\phi_{\text{dual 2}}}=\frac{64}{(a{\hat r}y_1)^2},\ B=0\,.
\eea
As expected, this geometry is a dual of (\ref{Group1lim}) under NATD along the two--dimensional sphere. Therefore, background (\ref{DualSU2pFlat}) can be obtained from the system (\ref{S3startFlat}) by performing a T--duality along the three--dimensional torus followed by a non--abelian T--duality along $S^2$. The Ramond--Ramond fields work as well. 

According to the results of the previous subsection, the geometry (\ref{DualSU2p}) is equivalent to dualization of (\ref{S3start}) along $S^3$ viewed as the $SO(4)/SO(3)$ coset. The large 
$\Lambda$ limit of this dual (\ref{MetrCosetDual}) gives
\bea\label{S3cosetFlat}
d{s}^2_{\text{coset dual}}=ds^2_{7}+\frac{dX_1^2}{a}+\bigg(\frac{dx_2}{x_3}\bigg)^2\frac{X_1^2}{a}+\bigg[dx_2\bigg(\frac{x_2}{x_3}\bigg)+dx_3\bigg]^2\frac{a}{X_1^2}
\,.
\eea
The change of variables (\ref{SU2map}) reduces to a simpler map between coordinates used in (\ref{DualSU2pFlat}) and (\ref{S3cosetFlat}):
\bea\label{SU2mapFlat}
X_1=\frac{{\hat r}}{2}\,,\quad\quad x_2=\frac{y_2}{4}\,,\quad\quad x_3=\frac{y_1}{4}\,.
\eea

\subsection{Duality along the diagonal $SU(2)$}

Let us go back to the background (\ref{S3start}) where three-sphere is viewed as a group manifold (\ref{OmegaAsSU2}). The action of the sigma model is invariant under $SU(2)_L\times SU(2)_R$ acting as (\ref{ActSU2LR}), and in the previous subsection we performed an NATD by gauging the action by $h_L$. In this subsection we will take an alternative path and perform a duality by gauging the action of the diagonal $SU(2)$ with $h_L=h_R^{-1}$ in  (\ref{ActSU2LR}).

To carry out the duality along the diagonal subgroup using the formalism reviewed in section \ref{SecReview}, we 
begin with rewriting the action describing the metric (\ref{S3start}), (\ref{OmegaAsSU2}) in terms of left invariant forms:
\bea\label{ActDiagSU2v1}
S=\int d^2\sigma\bigg[-E L^a_+L^a_-+g_{\mu\nu}\d_+Y^\mu \d_-Y^\nu\bigg]\,. 
\eea
The action of the diagonal $SU(2)$ can be viewed either as an $SO(3)$ rotation of index $a$ or as a special case of (\ref{ActSU2LR}):
\bea\label{ActSU2LRx}
g\rightarrow h g h^{-1}\,.
\eea
We will focus on the latter option. Gauging of this group introduces covariant derivatives 

\bea
\partial_\pm g\rightarrow  D_\pm g=\partial_\pm g-[A_\pm, g]\,,\quad A_\pm=A^a_\pm \sigma_a\,,
\eea
where $\sigma_a$ are Pauli matrices. Gauge transformations can be used to put $g$ in the diagonal form:
\bea\label{gDiagGauge}
g=\begin{bmatrix}
e^{i\theta}&0\\
0&e^{-i\theta}
\end{bmatrix}\,.
\eea
Note that, in contrast to the examples discussed earlier, the group degrees of freedom cannot be gauged away completely, and one field $\theta$ will remains in the dual model. Gauging the action (\ref{ActDiagSU2v1}) and adding Lagrange multipliers for the field strength, we find\footnote{To avoid unnecessary clutter, we drop the $Y$ terms.} 
\bea
S_L&=&\int d^2\sigma E\bigg[2\d_+\theta \d_-\theta+\sum_{a=1}^2\mbox{Tr}(t^ag^{-1}[A_+,g])\mbox{Tr}(t^ag^{-1}[A_-,g])-i\mbox{Tr}[vF_{-+}]\bigg]\nn
&=&\int d^2\sigma 2E\bigg[\d_+\theta \d_-\theta+4(A^1_-A^1_++A^2_-A^2_+)\sin^2{\theta}\bigg]\\
&&+\int d^2\sigma(i\partial_+v^aA^a_--i\partial_-v^bA^b_+-A^a_+f_{ab}A^b_-)\,.
\nonumber
\eea
Here the antisymmetric matrix $f_{ab}$ is defined as in (\ref{dualMij}):
\bea 
f_{ab}\equiv f_{ab}{}^cv_c\,.
\eea
Solving equations of motion for the gauge potentials,
\bea
A^a_\pm=\pm i(M^{-1})^{ab} \partial_\pm v^b,\quad 
M_{ab}=(8\sin^2{\theta}EP_{ba}+f_{ba}),\quad
P_{ab}=\mbox{diag}(1,1,0)\,.
\eea
we arrive at the dual action
\bea\label{SU2diagDual}
S_{dual\ 3}=\int d^2\sigma\bigg[2E(\d_+\theta \d_-\theta)+(\partial_+v^a)(M^{-1})^{ab}(\partial_-v^b)\bigg]\,.   
\eea	
Observing that the gauge (\ref{gDiagGauge}) is preserved by the diagonal $SU(2)$ rotations $h$, which form a little group for the element (\ref{gDiagGauge}), we conclude that the action (\ref{SU2diagDual}) should be viewed as dualization with respect to the coset $SU(2)/U(1)$ so one needs to fix the gauge for the Lagrange multipliers by setting $v_3=0$. This leads to the final result 
\bea\label{SU2diagDualfin}
S_{dual\ 3}=\int d^2\sigma\bigg[2E(\d_+\theta \d_-\theta)+\frac{1}{8E\sin^2{\theta}}(\partial_+v^i \partial_-v^i)\bigg]\,.    
\eea
In contrast to examples discussed in previous subsections, this geometry has a residual $U(1)$ symmetry.


\section{Spectral flow for geometries with products of spheres}
\label{SecSpecFlow}

In the previous section we analyzed a web of dualities for geometries containing a three dimensional sphere. As we saw, there were only two distinct ways to carry out dualities: one can dualize either along $SU(2)$ or along the $SO(4)/SO(3)$ coset. The situation becomes more interesting when there are two $S^3$ factors: in this case one can dualize along each of the spheres or along a shifted direction. The second option is analogous to a well--known feature of the abelian T--duality: in the presence of two $U(1)$ symmetries, one can dualize along any combination of the isometry directions. Specifically, if the metric has a form
\bea\label{MetrAbelT2}
ds^2=f(dy_1+g dy_2+A_m dx^m)^2+h(dy_2+B_m dx^m)^2+g_{mn}dx^mdx^n,
\eea
where coefficients are functions of $x_k$, then one can perform a duality along any direction in the $(y_1,y_2)$ plane. Periodic identification of the coordinates,
\bea\label{yPeriod}
(y_1,y_2)\sim (y_1,y_2)+2\pi(R_1,R_2)\,,
\eea
generates a lattice in this plane, and only dualities along the lattice directions constitute exact symmetries of string theory. Dualities along other directions can be viewed as a technique for generating new non--equivalent solutions, and this method, known as TsT transformation, has been extensively used in the past \cite{Melvin}. Extensions of this technique to the case of non--abelian dualities will be explored in the next section. To prepare for that discussion, in this section we will introduce several counterparts of the starting geometry (\ref{MetrAbelT2}) where the $(y_1,y_2)$ torus is replaced by a manifold with non--abelian symmetries, and we will discus some properties of the resulting supergravity solutions.

\subsection{General construction: geometries with $S^k\times S^k$} 
\label{SecSubSpekFlowSS}

Let us consider geometries which have two $S^k$ factors,
\bea\label{SNtimesSN}
ds^2=A\, d\Omega_k^2+B\, d{\tilde\Omega}_k^2+g_{mn}dx^mdx^n\,,
\eea
and introduce mixings between the spheres. 

For $k=1$ the situation is well known: the metric (\ref{SNtimesSN}) is a special case of the geometry (\ref{MetrAbelT2}) containing a two--dimensional torus, and shifting the coordinates $y_1$ as $y_1\rightarrow y_1+\gamma y_2$, one finds:
\bea\label{GeomTorusGam}
ds^2=A(dy_1+\gamma dy_2)^2+B dy_2^2+g_{mn}dx^mdx^n\,.
\eea
Parameter $\gamma$ corresponds to a diffeomorphism\footnote{Locally this is true for any $\gamma$, but generically the boundary conditions (\ref{yPeriod}) are violated.}. Assuming that the geometry (\ref{GeomTorusGam}) is supported by fluxes that do not contain the Kalb--Ramond field, we conclude that the T--duality along $y_1$ direction eliminates $\gamma$ dependence from the metric and moves it to the $B$ field with a trivial field strength. On the other hand, T-duality along $y_2$ direction leads to a $\gamma$--dependent geometry. To see this, we first rewrite (\ref{GeomTorusGam}) as
\bea\label{GeomTorusGamS}
ds^2=F\left(dy_2+\frac{\gamma A}{F}dy_1\right)^2+\frac{AB}{F}dy_1^2+g_{mn}dx^mdx^n,\quad F=B+\gamma^2 A.
\eea
Then the dual metric is
\bea\label{GeomTorusGamSD}
ds^2=\frac{1}{F}d{\tilde y}_2+\frac{AB}{F}dy_1^2+g_{mn}dx^mdx^n,
\eea
In this section we will explore higher--dimensional counterparts of shifts (\ref{GeomTorusGam})--(\ref{GeomTorusGamS}), and the non--abelian analogs of the dual (\ref{GeomTorusGamS}) will be studied in sections \ref{SecSfTdGroup} and \ref{SecSfTdCoset}.

To introduce a mixing between two spheres for general $k$ in (\ref{SNtimesSN}), we begin with writing each of the spheres in terms of $SO(k+1)/SO(k)$ cosets. Specifically, the metric of the first sphere can be written as
\bea
d\Omega_n^2=dy_j dy_j=V_k^T dg^T dg V_k,
\eea
where $(k+1)$ coordinates $y_j$ are subject to constraint $\sum y_jy_j=1$, which can be resolved by setting 
\bea
\left[\begin{array}{c}
y_1\\ y_2\\ \dots\\  y_{k+1}
\end{array}\right]=g V_n,\qquad
V_k=\left[\begin{array}{c}
1\\ 0\\ \dots\\  0
\end{array}\right]\,.
\eea
Here $g$ is the $(k+1)$--dimensional representation of the $SO(k+1)/SO(k)$ coset. Introducing a similar parameterization for the second sphere, we can rewrite the metric (\ref{SNtimesSN}) in terms of two cosets $(g,h)$ as 
\bea\label{SNtimesSNgh}
ds^2=A\, V_k^T dg^T dg V_k+B\, V_k^T dh^T dh V_k+g_{mn}dx^mdx^n\,,
\eea
The fluxes supporting the metric (\ref{SNtimesSN}) can be written in terms of $(g,h,V_k)$ as well, and some examples will be discussed below. To introduce a mixing between the spheres inspired by (\ref{GeomTorusGam}), we make a replacement 
\bea\label{SpecFlowGH}
h\rightarrow g^n h 
\eea
in (\ref{SNtimesSNgh}). To simplify the final answer and to use it in the later sections, we introduce the 
left invariant forms $(L^r,{\tilde L}^{{a}})$ as well as $n$--dependent generalizations 
${\tilde L}^{{a}(n)}$:
\bea\label{LformsOne}
L^r=-i\mbox{Tr}(t^rh^{-1}dh),\quad 
{\tilde L}^{{a}}=-i\mbox{Tr}(t^{{a}}g^{-1}dg),\quad
{\tilde L}^{{a}(n)}=-i\mbox{Tr}(t^{{a}}g^{-n}d(g^n))\,.
\eea
Recall that in this article the generators are normalized by 
\bea\label{GenerNorm}
\mbox{Tr}(t^r t^s)=\delta^{rs},\quad \mbox{Tr}(t^{{a}}t^{{b}})=\delta^{{{a}{b}}}\,.
\eea
In terms of the forms (\ref{LformsOne}), the metric (\ref{SNtimesSNgh}) is
\bea\label{SNtSNghFrm}
ds^2=-\left[A {\tilde L}^{{a}}{\tilde L}^{{b}} +B {L}^{{a}}{L}^{{b}}\right]V_k^T t^a t^b V_k+
g_{mn}dx^mdx^n\,,
\eea
and the replacement (\ref{SpecFlowGH}) corresponds to the substitution 
\bea\label{SecFlowForL}
L^r\rightarrow L^r-i\mbox{Tr}(t^rh^{-1}g^{-n}(dg^n)h)=L^r+{\tilde L}^{{a}(n)}{D}^{ar}\,.
\eea
in (\ref{SNtSNghFrm}). Here we defined 
\bea
{D}^{ab}=\mbox{Tr}(t^aht^bh^{-1}).\nonumber
\eea
To summarize, the transformed version of (\ref{SNtSNghFrm}) is 
\bea\label{SNtSNghFrmD}
ds^2=-\left[A {\tilde L}^{{a}}{\tilde L}^{{b}} +B 
({L}^{{a}}+{\tilde L}^{\tilde{c}(n)}{D}^{ca})
({L}^{{b}}+{\tilde L}^{\tilde{d}(n)}{D}^{db})\right]V_k^T t^a t^b V_k+
g_{mn}dx^mdx^n\,,
\eea
Although formally summation over $(a,b,c,d)$ indices in this expression is performed over $\frac{n(n+1)}{2}$ generators of $SO(n+1)$, only $n+n$ forms $({L}^{{r}},{\tilde L}^{{a}})$ are nontrivial since $h$ and $g$ parameterize cosets rather than groups. Expression (\ref{SNtSNghFrmD}) is a non--abelian counterpart of the metric (\ref{GeomTorusGam}).

Since the geometry (\ref{SNtSNghFrmD}) is obtained from (\ref{SNtimesSNgh}) by a change of coordinates, this metric has an $SO(k+1)\times SO(k+1)$ symmetry acting on the {\it original} variables used in (\ref{SNtSNghFrm}) as 
\bea\label{g0trans}
n=0:&&g_0\rightarrow M_g g_0,\quad h_0\rightarrow M_h h_0,\quad (M_g,M_h)\in SO(k+1)\,.
\eea
The variables after the replacement (\ref{SpecFlowGH}),
\bea
g=g_0,\quad h=(g_0)^{-n}h_0,
\eea
as well as the forms $(L^r,{\tilde L}^{\tilde{a}},{\tilde L}^{\tilde{a}(n)})$ constructed from them transform in a complicated way under (\ref{g0trans}). However, there are still residual symmetries with simple actions:
\bea\label{SymCoset1}
n=1:&&g\rightarrow M_g g N_h^{-1},\quad h\rightarrow N_h h,\quad M_g\in SO(k+1),\
N_h\in SO(k);\nn
n=-1:&&g\rightarrow M_g g,\quad h\rightarrow M_g h,\quad M_g\in SO(k+1);\\
n\ne 0,\pm 1:&& g\rightarrow N g N^{-1},\quad h\rightarrow N h,\quad N\in SO(k).\nonumber
\eea
For future reference, the explicit symmetries of the action (\ref{SNtSNghFrmD}) for different values of $n$ are summarized in Table \ref{SymCosetTable}.
\begin{table}
\begin{center}
\begin{tabular}{|c|ll|c|}
\hline
$n$&{transformation}&&{symmetry}\\
\hline
$n=0$&$g\rightarrow M_1 g$&$ h\rightarrow M_2 h$&$SO(k+1)\times SO(k+1)$\\
\hline
$n=1$&$g\rightarrow M g N^{-1}$&$h\rightarrow N h$&$SO(k+1)\times SO(k)$\\
\hline
$n=-1$&$g\rightarrow M g,$&$ h\rightarrow M h$&$SO(k+1)$\\
\hline
$n\ne 0,\pm1$&$g\rightarrow N g N^{-1},$&$ h\rightarrow N h$&$SO(k)$\\
\hline
\end{tabular}
\end{center}
\caption{Explicit symmetries of $n$--twisted products $\{[SO(k+1)/SO(k)]^2\}_n$.}
\label{SymCosetTable}
\end{table}
In section \ref{SecSfTdCoset} dualities along directions associated with these symmetries will be used to construct new geometries. 

The non--abelian counterpart of the rearranged metric (\ref{GeomTorusGamS}) for generic $n$ looks rather complicated since it involves both ${\tilde L}^{{a}}$ and ${\tilde L}^{{a}(n)}$. For $n=1$ this complication disappears and one gets
\bea\label{GTnabGamS}
ds^2=-\left[F ({\tilde L}^{{a}}+\frac{B}{F}D^{ae}L^e) ({\tilde L}^{{b}}+\frac{B}{F}D^{bf}L^f) +\frac{AB}{F}{L}^{{a}}
{L}^{{b}}\right]V_k^T t^a t^b V_k+
g_{mn}dx^mdx^n\,.
\eea
As in (\ref{GeomTorusGamS}) we defined
\bea
F=A+B. 
\eea
In the process of deriving (\ref{GTnabGamS}) we used two properties of matrix $D$:
\bea
D^{ab} D^{ac}=\delta^{bc},\quad 
{D}^{ca}{D}^{db}V_k^T t^a t^b V_k=V_k^T t^c t^d V_k\,.\nonumber
\eea
The dualities discussed in previous section can be easily applied to the product space (\ref{SNtimesSN}), and for $k=3$ the results are given by (\ref{MetrCosetDual}), (\ref{Group1}), (\ref{DualSU2p}), (\ref{SU2diagDual}) for each of the spheres. Dualization of (\ref{SNtSNghFrm}) along $h$ directions while keeping $g$ fixed is a non-abelian counterpart of dualizing (\ref{GeomTorusGam}) along $y_1$, and we will call this procedure a ``trivial duality'' for the twisted case. Dualization of (\ref{SNtSNghFrm}) along $g$ directions while keeping $h$ fixed leads a non--abelian counterpart of (\ref{GeomTorusGamSD}) that has a non--trivial $\gamma$ dependence, so we will call this procedure a ``nontrivial duality''\footnote{Note that in the non--abelian case, both ``trivial'' and ``nontrivial'' procedures lead to interesting dependence of the dual geometry on the mixing parameter $n$, so the labels inherited from the abelian situation are used only to distinguish the options.}. The relevant dualities will be discussed in section \ref{SecSfTdCoset}. 

\bigskip

While the transformation (\ref{SpecFlowGH}) leading to the mixture (\ref{GTnabGamS}) can be performed for spheres in arbitrary dimensions, for $k=3$ there is another option. As reviewed in section \ref{SecS3}, three--sphere can be viewed either as a coset or as a group manifold, and so far have we have discussed the mixing coming from the first option. Alternatively, starting from a group description 
\bea\label{S3timesS3g}
ds^2&=&-\frac{A}{4}\mbox{tr}(g^{-1}dgg^{-1}dg)-\frac{B}{4}\mbox{tr}(h^{-1}dhh^{-1}dh)+g_{mn}dx^mdx^n\,\nn
&=&A{\tilde L}^{{a}}{\tilde L}^{{a}}+B{L}^{{a}}{L}^{{a}}+g_{mn}dx^mdx^n\,,\quad (g,p)\in SU(2)
\eea
instead of (\ref{SNtimesSNgh}), we can perform the following mixing:
\bea\label{SpecFlowGH3}
h\rightarrow g^n h.
\eea
As in the coset case discussed earlier, this mixing makes some symmetries of the geometry (\ref{S3timesS3g}) somewhat implicit, and transformations with simple action on $(g,h)$ are summarized in table \ref{SymGroupTable}. This is a counterpart of the table \ref{SymCosetTable} for the model based on a group rather than a coset.
\begin{table}
\begin{center}
\begin{tabular}{|c|ll|c|}
\hline
$n$&{transformation}&&{symmetry}\\
\hline
$n=0$&$g\rightarrow M_1 g M_2$&$ h\rightarrow M_4 h M_3$&$[SU(2)]^4$\\
\hline
$n=1$&$g\rightarrow M_1 g M_2$&$h\rightarrow M^{-1}_2 h M_3$&$[SU(2)]^3$\\
\hline
$n=-1$&$g\rightarrow M_1 g M_2,$&$ h\rightarrow M_1 h M_3$&$[SU(2)]^3$\\
\hline
$n\ne 0,\pm1$&$g\rightarrow M_1 g M_1^{-1},$&$ h\rightarrow M_1 h M_2$&$[SU(2)]^2$\\
\hline
\end{tabular}
\end{center}
\caption{Explicit symmetries of the $n$--twisted product $\{SU(2)\times SU(2)\}_n$. Here both three--dimensional spheres are viewed as $SU(2)$ group manifolds.}
\label{SymGroupTable}
\end{table}

Rewriting (\ref{S3timesS3g}) in terms of left invariant forms (\ref{LformsOne}) and applying the transformation (\ref{SecFlowForL}), we find
\bea\label{S3specFlow}
ds^2&=&A {\tilde L}^{{a}}{\tilde L}^{{a}} +B 
({L}^{{a}}+{\tilde L}^{{c}(n)}{D}^{ca})
({L}^{{a}}+{\tilde L}^{{d}(n)}{D}^{da})+
g_{mn}dx^mdx^n\,\\
&=&A {\tilde L}^{{a}}{\tilde L}^{{a}} +B 
({R}^{{a}}+{\tilde L}^{{a}(n)})
({R}^{{a}}+{\tilde L}^{{a}(n)})+
g_{mn}dx^mdx^n\,.\nonumber
\eea
Here we defined the right--invariant forms $R^a=-i\mbox{Tr}(t^adh h^{-1})$, which make some of the symmetries (\ref{SymGroupTable}) more explicit.
For future reference we also write the counterpart of the rearranged expression (\ref{GTnabGamS}) with $n=1$ for the metric (\ref{S3specFlow}):
\bea\label{GTnabGamS3}
ds^2=F ({\tilde L}^{{a}}+\frac{B}{F}D^{ae}L^e) ({\tilde L}^{{a}}+\frac{B}{F}D^{af}L^f) +\frac{AB}{F}{L}^{{a}}
{L}^{{a}}+
g_{mn}dx^mdx^n\,.
\eea
Non--abelian T--dualities for the twisted geometries (\ref{GTnabGamS3}) will be studied in section \ref{SecSfTdGroup}, but before doing that it is useful to explore an intriguing analogy between (\ref{GTnabGamS3})--(\ref{S3timesS3g}) and the gravitational description of the spectral flow in the context of the AdS/CFT correspondence. Specifically, in the next subsection we will briefly review the spectral flow in $N=4$ superconformal field theories in two dimensions and its geometrical manifestation on the gravity side. Then in section \ref{SecSubSF10d} we will apply the twist (\ref{GTnabGamS3}) to regular geometries describing 1/2--BPS states in the $AdS_5/CFT_4$ correspondence and demonstrate that this twist can be viewed as a higher--dimensional counterpart of the spectral flow, and that the deformed metrics have very interesting algebraic structures.

\subsection{Spectral flow in $AdS_3\times S^3$}

Since we will be focusing on geometries (\ref{S3timesS3g}) arising in the AdS/CFT correspondence, it is useful to give a physical interpretation of the shift (\ref{SpecFlowGH3}) from the perspective of the dual field theory. To do so, we recall the situation for the Abelian case and its application to the AdS$_3$/CFT$_2$ correspondence. The dual of the AdS$_3\times$S$^3\times$X geometry
\bea\label{AdS3S3}
ds^2=L^2\left[-(1+\rho^2)dt^2+\frac{d\rho^2}{1+\rho^2}+\rho^2 d\chi^2+
d\theta^2+\cos^2\theta d\phi^2+\sin^2\theta d\psi^2\right]+ds_X^2
\eea
corresponds to the NS--NS vacuum in the dual CFT. The field theory side admits a symmetry known as spectral flow \cite{SchwSeib,SpecFlow}, which modifies the boundary conditions for fermions in the plane, so that 
\bea\label{SpecFlow}
\psi(e^{2\pi i}z)=-e^{-i\pi\eta}\psi(z)\,.
\eea
In particular, $\eta=0,2,4\dots$ describe states the NS--NS sector (including the vacuum dual to (\ref{AdS3S3})), and $\eta=1,3,5\dots$ cover the RR states. The field theoretic dual of the string theory on AdS$_3\times$S$^3\times$X has $N=4$ superconformal symmetry with an $SU(2)$ R charge, and spectral flow leads to the following transformation of bosonic generators:
\bea
L'_n=L_n-\eta J_n^3+\frac{c\eta^2}{24}\delta_{n,0},\quad (J^3)'_n=J_n^3-\frac{c\eta}{12}\delta_{n,0},\quad
(J^\pm)'_n=J^{\pm}_{n\pm\eta}\,.
\eea
On the gravity side, this flow corresponds to a change of variables in (\ref{AdS3S3}):
\bea\label{SpecFlowAdS3}
\phi\rightarrow \phi+\eta t,\quad \psi\rightarrow \psi+\eta \chi\,.
\eea
For $\eta=1$, the shifted version of the metric (\ref{AdS3S3}) becomes
\bea\label{AdS3S3sft}
ds^2&=&\frac{L^2}{W}\left[-(dt+Wc^2_\theta d{\tilde\phi})^2
+(d\chi-Ws_\theta^2 d{\tilde\psi})^2\right]\nn
&&+
L^2W\left[\frac{1}{W}\left(\frac{d\rho^2}{1+\rho^2}+d\theta^2\right)+
(\rho^2+1)c^2_\theta d\phi^2+\rho^2s^2_\theta d\psi^2\right]+ds_X^2\,.
\eea
Here 
\bea\label{AdS3W}
W=\frac{1}{\rho^2+\sin^2\theta}
\eea
is a harmonic function on the four dimensional flat base parameterized by coordinates 
$(\rho,\theta,{\tilde\phi},{\tilde\psi})$. 
The construction (\ref{AdS3S3sft}) can be extended to gravitational duals of all chiral primaries preserving half of supersymmetries \cite{fuzzD1D5}\footnote{To avoid unnecessary complications, we focus on geometries described by one harmonic function $Z$. The solution with two such functions, $(f_1,f_5)$, has a similar form.}:
\bea\label{D1D5twist}
ds^2&=&\frac{1}{W}\left[-(dt-A)^2+(dy+B)^2\right]+W(d{\bf x}\cdot d{\bf x})_4+ds_X^2,\nn
C_2&=&\frac{1}{W}(dt-A)\wedge (dy+B)+{\cal C}_2,\quad dB=-\star_4 dA,\quad d{\cal C}_2=-\star_4 dW.
\eea
Function $W$ satisfies the Laplace equation on the four--dimensional base, and gauge field $A$ satisfies the four--dimensional Maxwell's equations. 

Probe D1 branes (the twisted versions of the giant gravitons \cite{giant}) have a very simple description in this geometry. Imposing the static gauge and assuming fixed locations of branes on the base,
\bea
t=\tau,\quad y=\sigma,\quad x_i=x^{(0)}_i,
\eea
we arrive at the vanishing action:
\bea\label{DBIaDs3}
S=S_{DBI}+S_{CS}=T\int d\tau d\sigma \frac{1}{W}-T\int d\tau d\sigma \frac{1}{W}\,.
\eea
Variation of this action with respect to parameters $x^{(0)}_i$ demonstrates that the locations of branes on the base are indeed arbitrary. 

In the next subsection we will discuss a counterpart of the geometry (\ref{D1D5twist}) describing a twisted version of chiral primaries in $N=4$ SYM$_4$, and in section \ref{SecSubBranes} we will demonstrate that the twisted coordinates provide a very natural framework for describing probe branes on the bubbling geometries. 

\subsection{Bubbling geometries and spectral flow in 10d}
\label{SecSubSF10d}

Let us now go back to the geometries (\ref{S3timesS3g}). In the context of AdS/CFT correspondence a large class of such geometries describes the duals of chiral primaries in $N=4$ super--Yang--Mills \cite{LLM}, so it is natural to look for a counterpart of the spectral flow transformation (\ref{SpecFlowAdS3}). To get some intuition, we begin with writing the AdS$_5\times$S$^5$ geometry as (\ref{S3timesS3g}):
\bea\label{S3timesS3AdS5}
ds^2&=&\ell^2\left[-(1+\rho^2)dt^2+\frac{d\rho^2}{1+\rho^2}-
\frac{\rho^2}{4}\mbox{tr}(h^{-1}dhh^{-1}dh)\right]\nn
&&+\ell^2\left[d\theta^2+\cos^2\theta d\phi^2-
\frac{\sin^2\theta}{4}\mbox{tr}(g^{-1}dgg^{-1}dg)
\right]\,.
\eea
Comparing this with (\ref{AdS3S3}), we arrive at a natural counterpart of the spectral flow (\ref{SpecFlowAdS3}):
\bea\label{SpecFlowAdS5}
g\rightarrow h^kg,\quad \phi\rightarrow \phi+k t\,.
\eea
In particular, for $k=1$, the construction (\ref{GTnabGamS3}) leads to  a counterpart of the geometry (\ref{AdS3S3sft}):
\bea\label{AdS5shift}
ds^2&=&\frac{\ell^2}{W}\left\{(dt+Wc_\theta^2 d{\tilde\phi})^2-
({\tilde L}^{{a}}+Ws_\theta^2 D^{ae}L^e)^2\right\}\nn
&&+
\ell^2W\left[
\frac{1}{W}\left(\frac{d\rho^2}{1+\rho^2}+d\theta^2\right)+
(\rho^2+1)c^2_\theta d{\tilde\phi}^2+\rho^2s^2_\theta L^aL^a
\right]\,.
\eea
The expression in the square brackets parameterizes a flat six--dimensional space, and 
function $W$ is still given by (\ref{AdS3W}), but it is no longer harmonic. As we will see below, this function can be expresses in terms of two harmonic functions on the base.
Let us now extend the expression (\ref{AdS5shift}) to shifted versions of more general supergravity solutions known as bubbling geometries \cite{LLM}.

We begin with recalling the supergravity solutions dual to 1/2--BPS states in $N=4$ SYM. These geometries have two three-dimensional spheres, so they fall into the general category (\ref{SNtimesSN}) with $k=3$. 
The explicit solutions were found in \cite{LLM}:
\bea\label{BG metric}
ds^2
&=&-h^{-2}(dt+V_{i}dx^{i})^2+h^2(dy^2+dx^{i}dx^{i})+ye^Gd\Omega_3^2+ye^{-G}d\Tilde{\Omega}_3^2\,,\nn
F^{(5)}&=&F  \wedge d\Omega_3+\Tilde{F} \wedge d\tilde{\Omega}_3\,.
\eea
Various functions appearing in these expressions are given by
\bea
&& h^{-2}=2y\cosh{G},\quad z=\frac{1}{2}\tanh{G},\quad ydV=\star_3 dz\,,\nn
&&F=-\frac{1}{4}d[y^2 e^{2G}(dt+V)]-\frac{1}{4}y^3\star_3d\bigg[\frac{z+1/2}{y^2}\bigg],\\
&&{\tilde F}=-\frac{1}{4}d[y^2 e^{-2G}(dt+V)]-\frac{1}{4}y^3\star_3d\bigg[\frac{z-1/2}{y^2}\bigg]\,.\nonumber
\eea
The geometries are parameterized by one function $z$ satisfying a linear equation
\begin{equation}\label{eqnZ}
\partial_i\partial_i z+y\partial_y\bigg(\frac{\partial_y z}{y}\bigg)=0\,,  
\end{equation}
as well as some specific boundary conditions in the $y=0$ plane. 

Spectral flow (\ref{SpecFlowGH3}) 
by one unit leads to the geometry in the form (\ref{GTnabGamS3})
\bea\label{SpecFlowBubble}
ds^2
&=&\frac{1}{h^{2}}\left[-(dt+V)^2+ ({\tilde L}+wDL)^a ({\tilde L}+wDL)^a\right]+h^2\left[dy^2+
dx^{i}dx^{i}+y^2 d\Omega_3^2\right]\,,\nn
F_5&=&F\wedge (L+{\tilde L}^{(n)}D)^1\wedge (L+{\tilde L}^{(n)}D)^2\wedge (L+{\tilde L}^{(n)}D)^3+
\tilde{F}\wedge {\tilde L}^{\tilde{1}}\wedge {\tilde L}^{\tilde{2}}\wedge {\tilde L}^{\tilde{3}}
\,.
\eea
Here we introduced a convenient function $w$:
\bea
w=yh^2 e^G=\frac{e^G}{e^G+e^{-G}}=z+\frac{1}{2}\,.
\eea
The structure of the geometry (\ref{SpecFlowBubble}) is very similar to (\ref{D1D5twist}): it can be viewed as a fibration of $(t,{\tilde L})$ coordinates over a flat six--dimensional base. Furthermore, defining new functions 
\bea\label{ZpmDef}
Z_\pm=\frac{z\pm\frac{1}{2}}{y^2}\,,
\eea
we can rewrite the main dynamical equation (\ref{eqnZ}) as Laplace equations on the flat six--dimensional base:
\bea\label{LaplaceZZ}
\nabla_x^2 Z_\pm+\frac{1}{y^3}\d_y(y^3\d_y Z_\pm)=0.
\eea
Equation (\ref{LaplaceZZ}) is very intriguing, suggesting that it might be possible to extend solutions (\ref{SpecFlowBubble}) to geometries where $Z_\pm$ depend on all six coordinates. We leave 
investigation of this possibility to future work, and here we just write the expressions for $h$ and $V$ in terms of the harmonic functions and Hodge duality on the base:
\bea
h^{2}=\left[\frac{Z_+Z_-}{Z_++Z_-}\right]^{1/2},\quad
dV={y^2}\star_6\left[d(y^2 Z_\pm)\wedge d^3\Omega\right]\,.
\eea
In section \ref{SecSfTdGroup} we will discuss non--abelian T-dualities for the bubbling geometries in twisted coordinates (\ref{SpecFlowBubble}), but before doing that it is interesting to discuss probe branes in theses backgrounds. As we will see in the next subsection, the coordinates (\ref{SpecFlowBubble}) allow a very nice unified description of giant gravitons and dual giants.


\subsection{Probe branes in twisted coordinates}
\label{SecSubBranes}

Let us begin with a very brief review of giant gravitons \cite{giant} and dual giants \cite{giantDual} in the $AdS_5\times S^5$ background (\ref{S3timesS3AdS5}) supported by the five--form flux:
\bea\label{S3timesS3AdS5flux}
ds^2&=&\ell^2\left[-(1+\rho^2)dt^2+\frac{d\rho^2}{1+\rho^2}+
\rho^2 d{\tilde\Omega}_3^2
+d\theta^2+\cos^2\theta d\phi^2+
\sin^2\theta d\Omega_3^2
\right]\,,\nn
C_4&=&\frac{\ell^4}{4}\left[\rho^4 dt\wedge d^3{\tilde\Omega}+\sin^4\theta d\phi\wedge d^3\Omega\right]\,.
\eea
Dual giant gravitons are the spherical branes wrapping the sphere ${\tilde S}^3$ and rotating in the $\phi$ direction. Therefore the ansatz for its embedding in the geometry (\ref{S3timesS3AdS5flux}) is given by \cite{giantDual}
\bea
t=\tau,\quad {\tilde S}^3\rightarrow (\sigma_1,\sigma_2,\sigma_3),\quad 
\phi=\phi(t),\quad \theta=0,\quad \rho=\rho_0.
\eea
The action describing such branes is 
\bea
S=S_{DBI}+S_{CS}=-T\ell^4 \int d\tau d^3\sigma \rho_0^3\sqrt{\rho_0^2+1-{\dot\phi}^2}+
T\ell^4 \int d\tau d^3\sigma \rho_0^4\,.
\eea
Variations with respect to $\phi$ and $\rho_0$ imply that $\psi=\tau$ and parameter $\rho_0$ is arbitrary\footnote{There are additional solutions with $\phi=a \tau$ and specific values of $\rho_0$, but they correspond to maxima rather than minima of the action \cite{giantDual}.}. The original giant gravitons \cite{giant} have a similar description, but they are wrapping $S^3$ rather than ${\tilde S}^3$.

While the traditional description outlined above treats giants and dual giants separately, the twisted coordinates (\ref{AdS5shift}) allows one to view them on the same footing. Specifically, imposing the ansatz 
\bea\label{twistedGiants}
t=\tau,\quad {\tilde L}\rightarrow (\sigma_1,\sigma_2,\sigma_3),\quad 
\quad \theta=\theta_0,\quad \rho=\rho_0,\quad ({\tilde\phi},L^a) - \mbox{fixed}\,,
\eea
one finds the action for such D3 branes:
\bea
S=S_{DBI}+S_{CS}=-T\ell^4 \int d\tau d^3\sigma [\rho_0^2+\sin^2\theta_0]^2+
T\ell^4 \int d\tau d^3\sigma  [\rho_0^4+\sin^4\theta_0]\,.
\eea
In contrast to (\ref{DBIaDs3}), this action doesn't vanish for generic $(\rho_0,\theta_0)$. Variations with respect to $(\rho_0,\theta_0)$ lead to two branches of solutions describing giants and dual giants:
\bea
&&\mbox{giant gravitons}:\quad \rho_0=0,\quad \forall \theta_0;\nn
&&\mbox{dual giants}:\quad \theta_0=0,\quad \forall \rho_0.\nonumber
\eea 
Note that both branches are covered by the same ansatz (\ref{twistedGiants}), in contrast to the traditional description in the standard coordinates (\ref{S3timesS3AdS5flux}). 

The analysis presented above can be easily extended to the general geometries (\ref{SpecFlowBubble}). The counterpart of the ansatz (\ref{twistedGiants}) is
\bea\label{twistedGiantsA}
t=\tau,\quad {\tilde L}\rightarrow (\sigma_1,\sigma_2,\sigma_3),\quad 
(x_1,x_2,y,L^a) - \mbox{fixed},
\eea
and the action of the D3 brane becomes
\bea\label{ProbeD3bubl}
S=S_{DBI}+S_{CS}=-T \int d\tau d^3\sigma [h_0]^{-4}+
T \int d\tau d^3\sigma  [2y_0^2\cosh(2G_0)]\,.
\eea
Varying this action with respect to the parameter $y_0$ and recalling that 
$[h_0]^{-2}=2y_0\cosh G_0$, we find a constraint on $(G_0,y_0)$:
\bea
y_0\left[-(2\cosh G_0)^2+2\cosh(2G_0)\right]=0.
\eea
Assuming that $G_0=G(x_{10},x_{20},y_0)$ remains finite away from the $y=0$ plane\footnote{As shown in \cite{LLM}, this is one of the requirements for regularity of the geometry (\ref{BG metric}).}, we conclude that the branes must be located in the $y=0$ section of the six--dimensional base. This is consistent with well--known properties of (dual) giant gravitons in the bubbling geometries \cite{LLM}, but in contrast to the traditional approach, the description (\ref{SpecFlowBubble}), (\ref{twistedGiantsA}) treats both types of branes on the same footing. 

\bigskip

To summarize, in this section we have discussed a counterpart of the spectral flow for geometries containing two $k$--dimensional spheres. Furthermore, for $k=3$, when large classes of such supergravity solutions are known explicitly, we showed that the coordinates associated with spectral flow lead to a simple description of probe branes, where giant gravitons and dual giants are treated on the same footing. In the next section we will study the action of Non-Abelian T-Duality on the geometries with twisted spheres.

\section{NATD for geometries with twisted $S^3\times S^3$}
\label{SecSfTdGroup}

In this section we will explore applications of the spectral flow (\ref{SpecFlowAdS5}) followed by T--dualities along $g$ and $h$ to a subclass of geometries (\ref{S3timesS3g}) describing duals of chiral primaries in the $N=4$ SYM. This will generate a large class of new supergravity solutions describing field theories with non-standard boundary conditions for the fermions.

\begin{figure}
\[
\begin{tikzpicture}[scale=7]
\node (A) at (0,0) {$[S^3\times S^3]_n$};
\node (F) at (0.35,0.35) {A};
\node (C) at (0.35,-0.35) {B};
\node (D) at (0.7,0) {C};
\node (E) at (1.2,0) {D};
\draw[->] (A) -- (F) node[midway,left]{\scriptsize{Dual along $g$}};
\draw[->] (A) -- (C) node[midway,left]{\scriptsize{Dual along $h$}};
\draw[->] (F) -- (D) node[midway,right]{\scriptsize{Dual along $h$}};
\draw[->] (C) -- (D) node[midway,right]{\scriptsize{Dual along $g$}};
\draw[->] (D) -- (E) node[midway,below]{\scriptsize{Dual along $SO(3)/SO(2)$}};
\end{tikzpicture}
\]
\caption{Dualities for geometries with $S^3\times S^3$.}
\label{FigStmS}
\end{figure}
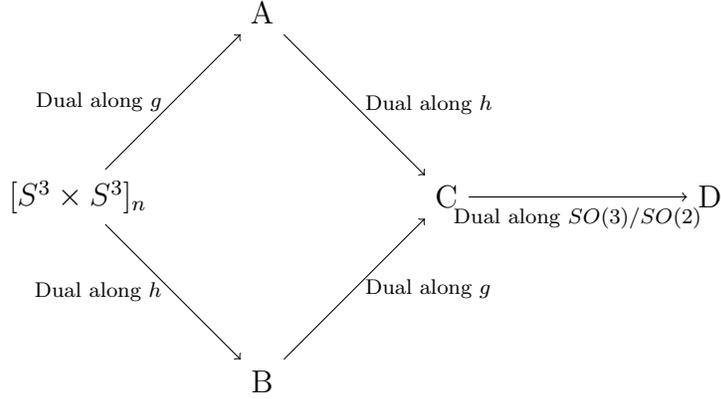

Although the original geometries (\ref{S3timesS3g}), (\ref{BG metric}) have 
$SO(4)\times SO(4)\simeq [SU(2)]^4$ symmetry, the rearranged metrics (\ref{GTnabGamS3}) and  (\ref{SpecFlowBubble}) make some of these $SU(2)$s implicit. As indicated in table (\ref{SymGroupTable}), for generic values of $n$, only the $[SU(2)]^2$ symmetries are explicit, and they act as the $SO(3)$ rotations of index $a$ in (\ref{S3timesS3g}) and as 
$h\rightarrow hM$. For $n=\pm 1$, there is an additional $SU(2)$ symmetry that acts on the group element $g$ either from the left of from the right. This leads to the following possibilities for the non-abelian T-dualities:
\label{ListOptionsDual}
\begin{enumerate}[A.]
\item T-duality along the $h$ direction. The abelian counterpart of this procedure, the dualization of (\ref{GeomTorusGam}) along the $y_1$ direction, leads to the geometry where information about the twist parameter $\gamma$ is lost, so we will denote this procedure as a ``duality along a trivial direction''. Note that this is just a label since in the non--abelian case the dependence on the spectral flow parameter $n$ will be highly nontrivial. The relevant geometries will be constructed in section \ref{SecSubOptA}.
\item T-duality along the $g$ direction that can be performed only for $n=0$ and $n=\pm 1$. The abelian counterpart of this procedure, the transition from (\ref{GeomTorusGamS}) to (\ref{GeomTorusGamSD}) leads to the geometry with non--trivial dependence on the twist parameter $\gamma$, so we will denote this procedure as a ``duality along a non--trivial direction''. The non--abelian version of the transition from  (\ref{GeomTorusGamS}) to (\ref{GeomTorusGamSD})  will be discussed in section \ref{SecSubOptB}.
\item Options A and B can be combined to perform a duality along both $h$ and $g$. The corresponding geometries will be constructed in section \ref{SecSubOptC}.
\item While the options A-C above exhaust all NATD along group manifolds that can be applied to (\ref{GTnabGamS3}) and (\ref{SpecFlowBubble}), the final results still have the $SO(3)$ isometry corresponding to rotation of index $a$. Dualization along this $SO(3)$ as a coset would lead to options A$'$, B$'$, C$'$. To avoid unnecessary clutter, in section \ref{SecSubOptD} we will focus only on the option C$'\equiv\,$D, and the other ones can be discussed in the same manner.
\end{enumerate}
The pictorial summary of the four options listed above is presented in figure \ref{FigStmS}. 

\subsection{Option A: duality along the ``trivial direction''}
\label{SecSubOptA}

In this subsection we will apply the spectral flow by $n$ units the background (\ref{S3timesS3g}) and dualize the resulting geometry along directions $h$. As discussed above, the abelian counterpart of such duality removed dependence on the spectral flow parameter, but this will not be true in the non--abelian case. The final metric and fluxes are given by (\ref{gBmatrT1}) and (\ref{RRaswOptA}). 

To perform the spectral flow and subsequent T--dualities for the background (\ref{S3timesS3g}), we begin with writing the bosonic part of the sigma model for this geometry in terms of the left invariant forms $(L_i^r,{\tilde L}_i^{\tilde{a}})$:
\bea\label{S3S3actL}
S=\int d^2\sigma [E(x)L^a_+L^a_-+F(x){\tilde L}^{\tilde{a}}_+
{\tilde L}^{\tilde{b}}_-+
g_{mn}\d_+ x^m\d_-x^n]\,.   
\eea
These forms are defined by\footnote{Recall that that generators $(t^r,t^{\tilde a})$ are normalized by (\ref{GenerNorm}).} 
\bea\label{Lforms}
L_\pm^r=-i\mbox{Tr}(t^rp^{-1}\partial_\pm p),\quad 
{\tilde L}_\pm^{\tilde{a}}=-i\mbox{Tr}(t^{\tilde{a}}g^{-1}\partial_\pm g).
\eea
The spectral flow is defined as a shift (\ref{SpecFlowGH}) combined with some change of coordinates $x^m$. This leads to a shift (\ref{SecFlowForL}) of the left invariant forms:
\bea\label{SecFlowForLaa}
L_\pm^r\rightarrow L_\pm^r+{\tilde L}_\pm^{\tilde{a}(n)}{D}^{ar}\,,\quad
{\tilde L}_\pm^{\tilde a}\rightarrow {\tilde L}_\pm^{\tilde a}
\eea 
The forms ${\tilde L}_\pm^{\tilde{a}(n)}$ are given by\footnote{These expressions are justified in the Appendix \ref{AppSubLn}.}
\bea\label{SpecFlowN}
{\tilde L}_\pm^{{a}(n)}\equiv-i\mbox{Tr}(t^{{a}}g^{-n}\partial_\pm g^n)={\tilde L}^{{b}}_\pm  \Bigg[
\sum_{s=0}^{n-1}{\tilde D}^s\Bigg]_{ba}\,,\quad 
{\tilde D}^{ab}=\mbox{Tr}(t^agt^bg^{-1}).
\eea
Replacements (\ref{SpecFlowN}) in (\ref{S3S3actL}) lead to the action
\bea\label{ActionSF}
S&=&\int d^2\sigma [E(L_+^r+{D_{a}}^r{\tilde L}_+^{{a}(n)})
(L_-^r+{D_{b}}^r{\tilde L}_-^{{b}(n)})+F{\tilde L}^{{a}}_+
{\tilde L}^{{a}}_-+
g_{mn}\d_+ x^m\d_-x^n]\,\\
&=&\int d^2\sigma [E(R_+^a+{\tilde L}_+^{{a}(n)})
(R_-^a+{\tilde L}_-^{a(n)})+F{\tilde L}^{{a}}_+
{\tilde L}^{{a}}_-+
g_{mn}\d_+ x^m\d_-x^n]\,.\nonumber
\eea
For example, (\ref{SpecFlowBubble}) describes a special case of a spectral flow of the bubbling geometries (\ref{BG metric}) by $n=1$ unit. The system (\ref{ActionSF}) can be dualized along the $p$ or $g$ directions, and before doing that let us make some comments about the Abelian case. 

The Abelian counterpart of the relation (\ref{SpecFlowN}) and the action (\ref{ActionSF}) are
\bea\label{ActionSFabel}
&&L^r\rightarrow L^r+n{\tilde L}^{r}=
L^r+{\tilde L}^{r(n)},\quad 
{\tilde L}^{\tilde a}\rightarrow {\tilde L}^{\tilde a},\quad
{\tilde D}^{ab}=\delta^{ab}\,,\nonumber\\
S&=&\int d^2\sigma [E(x)(\d z^r+n\d y^r)^2
+F(x)(\d y^a)^2+
g_{mn}\d_+ x^m\d_-x^n]\,, 
\eea
so dualization along $z$ directions leads to a metric that doesn't depend on $n$ and to a trivial $H$--field. In contrast to this, dualization along $y$ directions produces a nontrivial $n$--dependent Melvin twist \cite{Melvin}. In the non--abelian case (\ref{ActionSF}), both $p$ and $g$ lead to non--trivial $n$ dependence in the dual geometries. We begin with exploring the first option, and we will consider duality along $g$ in the next subsection. 

To perform a T--duality along the $h$ direction following the procedure outlined in section \ref{SecReview}, we gauge this $SU(2)$ symmetry by introducing covariant derivatives ($\partial_\pm h\rightarrow \partial_\pm h-hA_\pm$) and Lagrange multipliers $v_{\tilde{i}}$ for the gauge field $A_\pm$. Imposing the $h=\mathbb{I}$ gauge in the resulting action, we find\footnote{Here and below we use a shorthand notation ${\cal L}_\perp=g_{mn}\d_+x^n\d_-x^n$ term from the action.}
\bea
S&=&\int d^2\sigma \left[E{\tilde L}_+^{\tilde{a}(n)}{\tilde L}_-^{\tilde{a}(n)}+
F{\tilde L}^{\tilde{a}}_+
{\tilde L}^{\tilde{a}}_-+iA^a_+[E{\tilde L}_-^{(n)}-\d_-v]^a+
i[E{\tilde L}^{(n)}_++\partial_+v]^aA^a_-\right.\nonumber\\
&&\left.-A_+(E+f_v)A_-+{\cal L}_\perp\right]\,. \nonumber
\eea
Integrating out the gauge field, we arrive at the NATD dual of (\ref{ActionSF}):
\bea\label{BublTDtriv}
&&S_{\text{$t_1$}}=\int d^2\sigma \left[E{\tilde L}_+^{(n)}{\tilde L}_-^{(n)}+
F{\tilde L}_+{\tilde L}_-+({\tilde L}^{(n)}_+E+\partial_+v)\hat{M}^{-1}(\partial_-v-E{\tilde L}^{(n)}_-)+{\cal L}_\perp\right]\,.\nn
&&e^{-2\phi_{t_1}}=\mbox{det}(\hat{M})\,.
\eea
Here we defined some convenient quantities:
\bea
\hat{M}=(E+f^{(v)}),\quad {f^{(v)}}_{ab}=v_{{k}}{f_{ab}}^{{k}}\,.
\eea
For future reference we also give the expressions for the gauge field:
\bea
A_+&=i\hat{M}^{-T}({\tilde L}^{(n)}_+E+\partial_+v),\quad\quad A_-=-i\hat{M}^{-1}(\partial_-v-E{\tilde L}^{(n)}_-).
\eea
To write the background (\ref{BublTDtriv}) in a concise form, we define shifted derivatives $\mathscr{D}^{(n)}_\pm$ by
\bea\label{CovarDtriv}
\mathscr{D}^{(n)}_\pm v_{\tilde{i}}=\partial_\pm v_{\tilde{i}}+ 
\sqrt{2}\epsilon_{ijk}v_{\tilde{j}}{\tilde L}^{k(n)}_\pm\,.
\eea
Then the metric and the $B$ field become
\bea\label{SolnTdualTrivD}
ds^2&=&\frac{1}{E(E^2+2v_kv_k)}(\mathscr{D}^{(n)} v_{\tilde{i}})(E^2\delta_{ij}+2v_iv_j)(\mathscr{D}^{(n)} v_{\tilde{j}})+F{\tilde L}_+{\tilde L}_-+ds_\perp^2\,,\\
B&=&-\frac{\sqrt{2}}{E^2+2v^2}\eps_{ijk}v_i [\mathscr{D}^{(n)} v_{\tilde{j}}]\wedge 
[\mathscr{D}^{(n)} v_{\tilde{k}}]-2{\tilde L}^{(n)a}\wedge dv_a-{\sqrt{2}}\eps_{ijk}v_i
{\tilde L}^{(n)j}\wedge {\tilde L}^{(n)k}\,.\nonumber
\eea
Interestingly, the matrix $(g+B)^{-1}$ constructed from the metric and the Kalb--Ramond field $B$ appearing in (\ref{BublTDtriv}) satisfies a simple relation
\bea\label{gBmatrT1}
&&S_{\text{$t_1$}}=\int d^2\sigma \left\{[\d_+v, L_+^{(n)}](g+B)\left[
\begin{array}{c}
\d_-v\\ L_-^{(n)}
\end{array}\right]+{\cal L}_\perp\right\},\nn
&&(g+B)^{-1}=\left[\begin{array}{cc}
FV_nW_n+\sqrt{2}\eps_{ab{c}}v_{\tilde{c}}&V_n\\
-V_n&E^{-1}V_n
\end{array}
\right],\quad V_n=\left[I+\frac{F}{E}W_n\right]^{-1}\,.
\eea
The matrix $W$ is defined by a relation
\bea
F{\tilde L}_+{\tilde L}_-=F{\tilde L}^{(n)}_+W_n{\tilde L}_-^{(n)}\,,
\eea
and while its explicit form is rather complicated for a general $n$, in the most interesting case we find that
$W_1=1$.

Up to now we have focused only on the NS--NS fields for the geometry (\ref{S3S3actL}) and its dual. Let 
us now determine the Ramond--Ramond fluxes assuming that the original metric  (\ref{S3S3actL}) is supported only by the RR five--form. The symmetries of the problem imply that such flux must have the form 
\bea\label{Start5form}
F^{(5)}=G_2 \wedge d\Omega_3+\tilde{G}_2 \wedge d\tilde{\Omega}_3,
\eea
and specific examples of the system (\ref{S3S3actL}), (\ref{Start5form}) include so-called bubbling geometries (\ref{BG metric}). Application of the spectral flow (\ref{SpecFlowN}) to the five--form (\ref{Start5form}) and subsequent T--duality are performed in the Appendix \ref{AppRRopA}, and the results are given by (\ref{StrengthOptA}) and (\ref{PotOptA}). Here we just quote the answer (\ref{PotOptA}) for the derivatives of the RR potentials,
\bea\label{RRaswOptA}
&&d\t{C}_1=-\frac{G_2}{8},\quad d\t{C}_3=d\left[\frac{G_2}{8}\t{L}^{a(n)}v_a\right]\,,    \nn
&&d\t{C}_5=
\frac{G_2}{16}d\left[v_a\t{L}^{a(n)}\wedge (\t{L}^{b(n)}dv_b)\right]-\frac{\sqrt{2}\tilde{G}_2}{16}(\tilde{L}^{1}\tilde{L}^{2}\tilde{L}^{3})d(v_av_a)\,,\\
&&d\t{C}_7=\left[-\frac{\t{G}_2}{8}(\t{L}^1\t{L}^2\t{L}^3)-\frac{G_2}{8}(\t{L}^{1(n)}\t{L}^{3(n)}\t{L}^{3(n)})\right](dv_1dv_2dv_3) \,,
\nonumber
\eea
and recall the expressions for the fields strengths in terms of them:
\bea\label{ChernF246main}
&&\hat{F}_2=d\t{C}_1,\quad \hat{F}_4=d\t{C}_3+B\wedge \hat{F}_2,\quad
\hat{F}_6=d\t{C}_5+B\wedge \hat{F}_4-\frac{1}{2}B^2\wedge\hat{F}_2,\nn
&&\hat{F}_8=d\t{C}_7+B\wedge \hat{F}_6-\frac{1}{2}B^2\wedge \hat{F}_4+\frac{1}{6}B^3\wedge\hat{F}_2\,.
\eea
The explicit expressions for $(\hat{F}_2,\hat{F}_2,\hat{F}_2,\hat{F}_2)$ are given by (\ref{StrengthOptA}).

Let us now discuss D branes in the background (\ref{SolnTdualTrivD}), (\ref{RRaswOptA}). To compare with the analysis presented in section \ref{SecSubBranes} we will focus on $n=1$. All branes discussed in section \ref{SecSubBranes} were wrapping directions ${\tilde L}$ and stretching in time\footnote{Recall that in the standard coordinates (\ref{SecSubBranes}), giant gravitons wrap $S^3$ rather than ${\tilde S}^3$. As demonstrated in section \ref{SecSubBranes}, in the coordinates (\ref{SpecFlowBubble}) corresponding to a spectral flow by one unit, this translates into wrapping of the ${\tilde L}$ directions, putting the giant giant gravitons and dual giants on the same footing. Flows by $n\ne 1$ units don't admit this unified treatment of supersymmetric branes in bubbling geometries.}, so one expects that after dualities described here, the branes will be wrapping six spacial directions parameterized by $(L,{\tilde L})$. The DBI action for such branes reads
\bea\label{DBIsecA}
S_{DBI}=-T\int d^6\sigma e^{-\phi}\sqrt{-\mbox{det}(g+B)_{ind}}\,,
\eea
where $g_{ind}$ and $B_{ind}$ denote the induced metric and the Kalb--Ramond field. While the expressions for various components these fields given by (\ref{SolnTdualTrivD}) are rather complicated, the determinant can be easily extracted from (\ref{gBmatrT1}):
\bea
\mbox{det}(g+B)_{ind}=g_{tt}\frac{(E+F)^3}{E(E^2+2 v^2)}J^2,\ \mbox{where}\
J\equiv \mbox{det} J_{ab},\ J_{ab}=\left[-i\mbox{Tr}(t^ap^{-1}\partial_{\sigma_b} p)\right]\,.\nonumber
\eea
Note that while the $3\times 3$ matrix $J_{ab}$ depends on the explicit parameterization of $SU(2)$ by the brane coordinates $(\sigma_1,\sigma_2,\sigma_3)$, the DBI action (\ref{DBIsecA}) is invariant under reparameterizations of these three coordinates. For the remaining directions we chose the static gauge:
\bea
t=\sigma_0,\quad v_1=\sigma_4,\quad v_2=\sigma_5,\quad v_3=\sigma_6.
\eea
Recalling the expressions for various warp factors in the bubbling geometries (\ref{BG metric}), and assuming that the branes are located at fixed values of $(x_1,x_2,y)$ coordinates, we find
\bea
E=\frac{y_0}{4}e^{G_0},\quad F=\frac{y_0}{4}ye^{-G_0},\quad g_{tt}=-h_0^{-2}=-y_0(e^{G_0}+e^{-G_0})\,.
\eea
Substitution into (\ref{DBIsecA}) gives
\bea
S_{DBI}=-T\int d^6\sigma\frac{1}{h_0} (E+F)^{3/2} J=-T\int \left\{\frac{1}{8}d^6\sigma J\right\} [h_0]^{-4}\,.
\eea
Apart from the expected change in the measure that appears in the curly brackets, this action coincides with the DBI contribution to (\ref{ProbeD3bubl}). 

\subsection{Option B: duality along the ``non--trivial direction''}
\label{SecSubOptB}

Let us go back to the action (\ref{ActionSF}) and perform the non--Abelian T--duality along the $SU(2)$ parameterized by $g$. In the abelian counterpart of the action (\ref{ActionSFabel}) this corresponds to the duality along $y$ direction, and such procedure leads to the Melvin twist \cite{Melvin} with a nontrivial dependence on $n$ in the metric, the Kalb--Ramond field, and the dilaton. For this reason we use the term ``non--trivial case'' even in the non--abelian situation to distinguish this procedure from NATD along $h$ discussed in section \ref{SecSubOptA}\footnote{Recall that the abelian counterpart of the operation discussed there is the dualization along $z$ directions in (\ref{ActionSFabel}) which leads to a trivial $n$--dependence in the metric, the dilaton and the $H$ field.}. 

In contrast to $SU(2)$ symmetry utilized in the last subsection, the symmetries associated with $g$ are present only for some values of $n$. In the notation of table \ref{SymGroupTable}, we will dualize along the symmetries associated with $M_1$ for $n=(0,1)$ and along the directions associated with $M_2$ for $n=-1$. In the former cases, gauging of the symmetry and gauge fixing amount to replacements
 \bea\label{DualGreplaceN01}
 {\tilde L}_\pm\rightarrow -i\mbox{Tr}[t^{\tilde{a}}g^{-1}(\d_\pm-A_\pm)g]\rightarrow 
 i\mbox{Tr}[t^{\tilde{a}}A_\pm],\quad 
 {\tilde L}_\pm^{(1)}={\tilde L}_\pm\rightarrow i\mbox{Tr}[t^{\tilde{a}}A_\pm]\,.
\eea
in the action (\ref{ActionSF}). 

For $n=-1$, we begin with rewriting the action (\ref{ActionSF}) in terms of the right--invariant forms $(R,{\tilde R})$,
\bea\label{ActionSFright}
S_{n=-1}=\int d^2\sigma [E(R_+^a-{\tilde R}_+^{{a}})
(R_-^a-{\tilde R}_-^{a})+F{\tilde R}^{{a}}_+
{\tilde R}^{{a}}_-+
g_{mn}\d_+ x^m\d_-x^n]\,.
\eea
Here we used the relation\footnote{Recall the definition (\ref{SpecFlowN}) of the forms 
${\tilde L}^{a(n)}$.}
\bea
{\tilde L}_\pm^{a(-1)}=-i\mbox{Tr}(t^{\tilde{a}}g{}\partial_\pm g^{-1})=
i\mbox{Tr}(t^{\tilde{a}}(\partial_\pm g) g^{-1})=-{\tilde R}^{a}_\pm\,.\nonumber
\eea
Gauging of the symmetry $g\rightarrow g M$ in (\ref{ActionSFright}) and subsequent gauge fixing to $g=I$ leads to the following replacements in (\ref{ActionSFright}):
\bea
{\tilde R}_\pm\rightarrow -i\mbox{Tr}[t^{\tilde{a}}(\d_\pm g-gA_\pm)g^{-1}]\rightarrow 
 i\mbox{Tr}[t^{\tilde{a}}A_\pm]\,.
\eea
Interestingly, after the gauge fixing, the actions for $n=(0,\pm1)$ are covered by the same simple formula:
\bea\label{ActionSFAfield}
S=\int d^2\sigma [E(R_+^a+in{\tilde A}_+^{{a}})
(R_-^a+in{\tilde A}_-^{a})-F{\tilde A}^{{a}}_+
{\tilde A}^{{a}}_-+
g_{mn}\d_+ x^m\d_-x^n]\,.
\eea
Adding the $[-i\int d^2\sigma \operatorname{Tr}(sF_{+-})]$ term to (\ref{ActionSFAfield}) 
and integrating out the gauge field, we find the NS--NS fields for the dual geometry:
\bea\label{OptBdualNS}
&&S_{\text{${nt}_1$}}=\int d^2\sigma \Big[EL^{{a}}_+L^{{b}}_-+(nE{D^b}_rL_+^r+\partial_+s^{b})(\hat{N}^{-1})_{bc}(\partial_-s^c-nE{D^c}_rL_-^r)+g_{mn}dx^ndx^n\Big]\,, \nn
&&e^{-2\phi_{\text{stage ${nt}_1$}}}=\operatorname{det}(\hat{N})\,.
\eea
Here we defined a convenient $3\times 3$ matrix ${\hat N}$:
\bea
\hat{N}^{}=(n^2E+F+f^{(s)}),\quad {f^{(s)}}_{ab}=s_{{k}}{f_{ab}}^{{k}}\,.
\eea
The metric and the $B$ field can be combined into a simple matrix similar to the one introduced in 
(\ref{gBmatrT1}):
\bea\label{gBmatrT2a}
S_{{nt_1}}=\int d^2\sigma [ K_+,\d_+s](g+B)\left[
\begin{array}{c}
K_-\\ \d_-s
\end{array}\right],\quad
(g+B)^{-1}=\left[\begin{array}{cc}
E^{-1}&-n I\\
n I&F+\sqrt{2}\eps_{ab{c}}s_{{c}}
\end{array}
\right]\,.
\eea
Here $K^a_\pm={D^a}_rL_\pm^r$. Note that here and in the rest of this subsection, the spectral parameter $n$ can take only one of three values:
\bea
n=0,1,-1\,.
\eea
For other values of $n$, the symmetries utilized in the dualization procedure leading to (\ref{OptBdualNS}) mix variables $g$ and $h$ making these group coordinates unsuitable for describing the dual solution and destroying the procedure outlined above. 

To rewrite the metric and the $B$--field in a compact form, it is convenient to introduce counterparts of the covariant derivatives (\ref{CovarDtriv}):
\bea\label{CovarDnontr}
\mathscr{D}^{(n)}_\pm s_{{i}}=\partial_\pm s_{{i}}+ 
\frac{\sqrt{2}En}{F+En^2}\epsilon_{ijk}s_{{j}}{K}^{k}_\pm\,,\quad
H_n=F+En^2.
\eea
Then the metric and the $B$ field become
\bea\label{SolnTlNonTrivD}
ds^2&=&\frac{H_n}{(H_n)^2+2s^2}(\mathscr{D}^{(n)} s_{{i}})\left[\delta_{ij}+\frac{2s_is_j}{[H_n]^2}\right](\mathscr{D}^{(n)} s_{\tilde{j}})+\frac{EF}{H_n}{L}^a{L}^a+g_{mn}dx^ndx^n\,,\\
B&=&-\frac{\sqrt{2}}{2[[H_n]^2+2s^2]}\eps_{ijk}s_i [\mathscr{D}^{(n)} s_{{j}}]\wedge 
[\mathscr{D}^{(n)} s_{{k}}]-\frac{nE}{H_n}{K}^{i}\wedge ds_i-
\frac{\sqrt{2}(nE)^2}{2[H_n]^2}\eps_{ijk}s_i
{K}^{j}\wedge {K}^{k}\nonumber
\eea
In contrast to (\ref{SolnTdualTrivD}), where $n$ dependence appears only through the shifted derivatives, the system (\ref{SolnTlNonTrivD}) has an explicit $n$ dependent coefficients as well. This is expected from the intuition from the Abelian case (\ref{ActionSFabel}): the ``trivial'' T--duality along $z$ directions leads to an $n$--independent geometry, while the duality along $y$ leads to an $n$--dependent Melvin twist. Specifically, dualization of the system (\ref{ActionSFabel}) along the $y$ directions gives
\bea
ds^2=\frac{1}{H_n}(d{\tilde y}^a)^2+\frac{EF}{H_n}(dz^r)^2+g_{mn}dx^ndx^n,\quad
B=\frac{nE}{H_n}dz^r \wedge d{\tilde y}^r\,.\nonumber
\eea
These expressions are reproduced by (\ref{SolnTlNonTrivD}) in the decompactification limit, which is accomplished by making rescalings
\bea
(E,F)\rightarrow \frac{1}{\eps^2} (E,F),\quad L^a\rightarrow \eps dz^a,\quad s_i\rightarrow 
\frac{{\tilde y}^i}{{\eps}}\,,\nonumber
\eea
and sending $\eps$ to zero. In particular, in this limit the covariant derivatives and  (\ref{CovarDnontr}) reduce to partial ones, and $K^a$ coincide with $L^a$. Similarly, the ``trivial'', i.e., $n$--independent T--dual of (\ref{ActionSFabel}) along $z$ directions is recovered from (\ref{SolnTdualTrivD}) by making rescalings 
\bea
(E,F)\rightarrow \frac{1}{\eps^2} (E,F),\quad {\tilde L}^a\rightarrow \eps dy^a,\quad v_i\rightarrow 
\frac{{\tilde z}^i}{{\eps}}\,,\nonumber
\eea
and sending $\eps$ to zero. In this limit the covariant derivatives  (\ref{CovarDtriv}) reduce to the partial ones. 

To determine the Ramond--Ramond fluxes supporting the geometry (\ref{gBmatrT2a}), we  assume that the starting point of the duality procedure (\ref{S3S3actL}) has the fluxes (\ref{Start5form}) twisted by the spectral flow (\ref{SpecFlowN}). Then application of the T--duality leads to the fluxes (\ref{StenOptB}), (\ref{StenOptBe2}), (\ref{StenOptBe3}). The detailed construction is presented in the Appendix \ref{AppRRoptB}, and here we quote only the derivatives of the RR potentials:
\bea\label{RRpotOptB}
&&d\t{C}_1=-\frac{1}{8}(\t{G}_2+nG_2),\quad
d\t{C}_3=-d\left[\frac{nG_2}{8}(L^as_a)\right],\nn
&&d\t{C}_5=d\left[\frac{nG_2}{2\times 8}(L^as_a)(L^ids_i)\right],\quad
d\t{C}_7=-\frac{G_2}{8}(L_1L_2L_3)(ds_1ds_2ds_3)\,.
\eea
The fields strengths (\ref{StenOptB}), (\ref{StenOptBe2}) can be recovered using the general relations (\ref{ChernF246main}).

As we discussed in section \ref{SecSubBranes}, probe D3 branes in the geometry (\ref{SpecFlowBubble}) wrap the directions parameterized by ${\tilde L}$ while being located at a point in other spacial coordinates. Under the non--abelian duality described here, such branes should be mapped into D0 stretching only in the time direction. Assuming that all other coordinates of such zero branes are fixed, we find the induced metric and a pullback of the RR field in the dual geometry (\ref{OptBdualNS}):
\bea
ds_2=g_{tt}d\tau^2,\quad e^{-2\phi_B}=\operatorname{det}(n^2E+F+f^{(s)}),\quad
P[C^{(1)}]=\left[nC_t+{\tilde C}_t\right]d\tau\,.
\eea  
Substituting the fields for the geometries (\ref{SpecFlowBubble}),
\bea
E=\frac{y_0}{4}e^{G_0},\quad F=\frac{y_0}{4}ye^{-G_0},\quad C_t=-\frac{1}{4}(y_0e^{G_0})^2,\quad
{\tilde C}_t=-\frac{1}{4}(y_0e^{-G_0})^2\,,\quad g_{tt}=-h_0^{-2}\,,\nonumber
\eea
we arrive at the action for the probe D0 branes:
\bea
S=S_{DBI}+S_{CS}=-T \int d\tau e^{-\phi_B} [h_0]^{-1}+
T \int d\tau  y_0^2\left[e^{-2G_0}+n e^{2G_0}\right]\,.
\eea
For $n=1$ and $s_i=0$, this action differs from (\ref{ProbeD3bubl}) only by the number of integrations. As in the discussion after equation (\ref{ProbeD3bubl}), variation with respect to $y_0$ implies that the branes can be located at arbitrary points $(x_{1},x_{2})_0$ in the $y=0$ plane, but now we have an additional restriction on the $s$ coordinates: all of them must vanish. This constraint comes from the $f^{(s)}$ term in the dilaton, and it is related to the non--abelian nature of duality: recall that in the absence of structure constants, dualilization of D3 branes along $T^3$ leads to D0 branes which can be located at arbitrary points on the dual torus.

\subsection{Option C: application of two T--dualities} 
\label{SecSubOptC}

In the last two subsections we have dualized the geometries (\ref{ActionSF}) along $g$ and $h$ directions. Let us now combine these two dualities. Specifically, we start with the action (\ref{BublTDtriv}) and perform an NATD along this isometry associated with $g$ directions. 
 
Following the procedure reviewed in section \ref{SecReview}, we gauge the $SU(2)$ symmetry and fix the gauge by imposing the condition $g=\mathbb{I}$. These two steps lead to the replacements  (\ref{DualGreplaceN01}) in  (\ref{BublTDtriv}). 
Introducing the Lagrange multipliers, and integrating out the gauge field, we arrive at the T--dual of the geometry (\ref{BublTDtriv}):
\bea\label{BublTTact}
&&\hskip -1.5cm 
S_{C}=\int d^2\sigma\big[(\partial_+s-n(\partial_+v)\hat{M}^{-1}E)\hat{Y}^{-1}(\partial_-s-nE\hat{M}^{-1}(\partial_-v))
+(\partial_+v)\hat{M}^{-1}(\partial_-v)\big]\,,\nn
&&\hskip -1.5cm 
e^{-2\phi_{C}}=\operatorname{det}(\hat{M})\operatorname{det}(\hat{Y})\,.
\eea
Here we denoted the Lagrange multipliers introduced by the second NATD by $s_i$ and defined convenient matrices
\begin{equation}
\hat{Y}=\big[(F+n^2E)-n^2E\hat{M}^{-1}E+f^{(s)}\big],\quad f^{(s)}_{ab}=s_{{k}}f_{ab}^{{k}},\quad\hat{M}=E+f^{(v)} \,.
\end{equation}
Once again, we observe that the metric and the B-field encoded by (\ref{BublTTact}) give rise to a simple inverse of the $(g+B)$ matrix in the $(v,s)$ basis\footnote{Recall that the duality along $g$ direction can be performed only for $n=0,\pm1$, so the expression (\ref{stage two trivial bg}) is applicable only to these cases.}:
\bea\label{stage two trivial bg}
(g+B)^{-1}=\left[\begin{array}{cc}
E+\sqrt{2}\eps_{abc}v_{\tilde{c}}&nE\\
nE&F+n^2E+\sqrt{2}\eps_{abc}s_c
\end{array}
\right]\,.
\eea
As expected, for $n=0$ (i.e., in the absence of the spectral flow), the last expression describes separate NATDs along two spheres. Note that an alternative approach to option C, starting with (\ref{SolnTlNonTrivD}) and applying a duality along directions $h$, leads to the same answers (\ref{BublTTact})--(\ref{stage two trivial bg}). This explicit calculation verifies the expected commutativity of the diagram shown in figure \ref{FigStmS}. 
The Ramond--Ramond fields supporting the geometry (\ref{BublTTact})--(\ref{stage two trivial bg}) are constructed in Appendix \ref{AppRRoptC}, and the final results (\ref{flux Gs A-1}) are not very illuminating.

\subsection{Option D: application of three T--dualities}
\label{SecSubOptD}

The geometry (\ref{BublTTact}) obtained in the previous subsection has an $SO(3)$ symmetry acting on vectors $s_i$ and $v_i$. Here we will perform a T--duality along this symmetry and generate a solution that has no isometries. The final result would correspond to option D listed on page \pageref{ListOptionsDual}.

The geometry (\ref{BublTTact}) is invariant under a simultaneous $SO(3)$ rotation of the $(s_i,v_i)$ coordinates,
\bea
s_i\rightarrow g_{ij}s_j,\quad v_i\rightarrow g_{ij}v_j\,.
\eea
To perform a duality along this $SO(3)$, we write the vectors $(s,v)$ as 
\bea
\left[\begin{array}{c}
v_1\\ v_2\\ v_3
\end{array}\right]=g_D\left[\begin{array}{c}
v\\ 0\\0
\end{array}\right]\equiv g_D{\bar v},\quad
\left[\begin{array}{c}
s_1\\ s_2\\ s_3
\end{array}\right]=g_D\left[\begin{array}{c}
s\cos\zeta \\ s\sin\zeta\\0
\end{array}\right]\equiv g_D{\bar s}\,.
\eea
where columns $({\bar v},{\bar s})$ are defined by
\bea\label{vBarParam}
{\bar v}=vV_1,\quad {\bar s}=sh_D V_1,\quad V_1=\left[\begin{array}{c}
s\\ 0\\ 0
\end{array}\right],\quad 
h_D=\left[\begin{array}{ccc}
\cos\zeta&-\sin\zeta&0\\ \sin\zeta&\cos\zeta&0\\ 0&0&1
\end{array}\right]\,.
\eea
Writing the matrix (\ref{stage two trivial bg}) in terms of these variables,
\bea\label{gBoptD}
(g+B)^{-1}=
\left[\begin{array}{cc}
g&0\\
0&g
\end{array}
\right]
\left[\begin{array}{cc}
E+\sqrt{2}\eps_{abc}{\bar v}_{\tilde{c}}&nE\\
nE&F+n^2E+\sqrt{2}\eps_{abc}{\bar s}_c
\end{array}
\right]
\left[\begin{array}{cc}
g^T&0\\
0&g^T
\end{array}
\right]=G M G^T\,,\nonumber\\
\eea
and substituting the result into the action, we find
\bea
S_{{C}}&=&\int d^2\sigma \Big[
(\d_+ v_i,\d_+ s_i)GM^{-1}G^T\left[\begin{array}{c}
\d_- v_i\\
\d_- s_i
\end{array}\right]
\Big]\\
&=&
\int d^2\sigma 
\left[\begin{array}{c}
L^{b}_+t^b {\bar v}+\frac{\bar v}{v}\d_+ v\\
(L^{b}_++R^{(b)h}_+)t^b {\bar s}+\frac{\bar s}{s}\d_+ s
\end{array}\right]^T
M^{-1}\left[\begin{array}{c}
L^{a}_-t^a {\bar v}+\frac{\bar v}{v}\d_- v\\
(L^{a}_-+R^{a(h)}_-)t^a {\bar s}+\frac{\bar s}{s}\d_- s
\end{array}\right]\,.\nonumber
\eea
To arrive at this expression we used the relations
\bea
g^{-1}\d_-(gh)=L_-^{a}t^a h+\d_- h=(L^{a}_-+R^{a(h)}_-)t^a h,\quad
(g^{-1}\d_+(gh))^T=-h^T(L^{a}_++R^{a(h)}_+)t^a\,.\nonumber
\eea
Duality along $g_D$ direction gives \bea\label{ActOptD}
S_{D}&=&
= \int d^2 \sigma W^a_+(X^{-1})_{ab} W^b_-
+\begin{bmatrix}
\frac{\bar{v}}{v}\partial_+v\\
\frac{\bar{s}}{s}\partial_+s+R^a_+t^a\bar{s}
\end{bmatrix}^TM^{-1}\begin{bmatrix}
\frac{\bar{v}}{v}\partial_-v\\
\frac{\bar{s}}{s}\partial_-s+R^a_-t^a\bar{s}
\end{bmatrix}.
\eea
The ingredients of this expression are defined by 
\bea
W_+&=&\partial_+w^a+\begin{bmatrix}
\frac{\bar{v}}{v}\partial_+v\\
R^d_+t^d\bar{s}+\frac{\bar{s}}{s}\partial_+s
\end{bmatrix}^TM^{-1}\begin{bmatrix}
t^a\bar{v}\\
t^a\bar{s}
\end{bmatrix}\,,\nn
W_-&=&\partial_-w^a-\begin{bmatrix}
\frac{\bar{v}}{v}\partial_-v\\
R^d_-t^d\bar{s}+\frac{\bar{s}}{s}\partial_-s
\end{bmatrix}^T(M^T)^{-1}\begin{bmatrix}
t^a\bar{v}\\
t^a\bar{s}
\end{bmatrix}\,,\\
X_{ab}&=&(f_w)_{ab}+\left[\begin{array}{c}
{\bar v}\\
{\bar s}
\end{array}\right]^T
t^bM^{-1}t^a\left[\begin{array}{c}
{\bar v}\\
{\bar s}
\end{array}\right]\,.\nonumber
\eea
To get some intuition about the action (\ref{ActOptD}), it is instructive to analyze $X_{ab}$, for $n=0$, when it is relatively simple since matrix $M$ in (\ref{gBoptD}) has a block--diagonal form. Inversion of this matrix leads to the expression for $X$,
\bea
(X_{ab})_{n=0}=(f_w)_{ab}+
\frac{1}{E^2+2{\bar v}^2}{\bar v}^T\left[Et^b t^a -
2i{\bar v}^c t^b t^c t^a\right]{\bar v}+
\frac{1}{F^2+2{\bar s}^2}
{\bar s}^T\left[Ft^b t^a -
2i{\bar s}^c t^b t^c t^a\right]{\bar s}\nonumber
\eea
Inversion of this matrix give the components of the metric and the Kalb--Ramond field in the 
$w^a$ directions. The symmetric part of the matrix $(X_{ab})_{n=0}$ is
\bea
(X_{ab})^{sym}_{n=0}=
\frac{E}{E^2+2{\bar v}^2}{\bar v}^T t^b t^a{\bar v}+
\frac{F}{F^2+2{\bar s}^2}
{\bar s}^Tt^b t^a{\bar s}\equiv
A_E {\bar v}^T t^b t^a{\bar v}+A_F {\bar s}^Tt^b t^a{\bar s}.
\nonumber
\eea
Recalling the definition (\ref{vBarParam}) of columns ${\bar v}$ and ${\bar s}$, we conclude that matrices ${\bar v}^T t^b t^a{\bar v}$ and ${\bar s}^T t^b t^a{\bar s}$ are degenerate, 
but $(X_{ab})^{sym}_{n=0}$ has a non-zero determinant unless $\zeta$ vanishes:
\bea\label{DetXzeta}
\mbox{det}[(X_{ab})^{sym}_{n=0}]=
-\frac{1}{8}A_EA_F(A_E+A_F)\sin^2\zeta.
\eea
This agrees with the general intuition that each of the cosets parameterized by columns (\ref{vBarParam}) leads to a degenerate matrix ${\bar v}^T t^b t^a{\bar v}$ or 
${\bar s}^Tt^b t^a{\bar s}$, while combination of these cosets recovers a three--dimensional orbit of $SO(3)$. This property persists for other values of $n$, and it is a consequence of a simple geometry: any vector in $R^3$ is left invariant under its little group $SO(2)$, while a system of two linearly independent vectors has trivial little group. This difference will become important in the next section when we will discuss spectral flow and non--abelian T-dualities along cosets. 

\section{NATD for twisted $S^k\times S^k$ viewed as cosets}
\label{SecSfTdCoset}

In the previous section we explored non--abelian T--dualities for twists of geometries with $SO(4)\times SO(4)$ symmetries, and our discussion was based on viewing the three--spheres before the twist as group manifold. Unfortunately this construction cannot be extended to spaces with $S^k\times S^k$ for other values of $k$, but an alternative form of the twist (spectral flow) for general $k$ was introduced in section \ref{SecSubSpekFlowSS}. Specifically, there we viewed the spheres as cosets and used their symmetries to construct the twisted action (\ref{SNtSNghFrmD}). The explicit symmetries of such actions are summarized in Table \ref{SymCosetTable}, and in this section we will perform non--abelian T--dualities along these symmetries. To be specific and to avoid unnecessary clutter, we will mostly focus on $S^3$, but qualitative features of our discussion apply to all $SO(k+1)/SO(k)$ cosets.

Let us consider the sigma model corresponding to the twisted metric (\ref{SNtSNghFrmD}):
\bea\label{CosetTwist1}
S=-\int d^2\sigma\left[A {\tilde L}_+^{{a}}{\tilde L}^{{b}}_- +B 
({L}^{{a}}_++{\tilde L}^{\tilde{c}(n)}_+{D}^{ca})
({L}^{{b}}_-+{\tilde L}^{\tilde{d}(n)}_-{D}^{db})\right]V_k^T t^a t^b V_k\,,
\eea
Here and below we keep only the terms containing left invariant forms and drop the transverse part of the metric which is irrelevant for our discussion. To perform T--duality along $g$ directions, it is convenient to rewrite (\ref{CosetTwist1}) as 
\bea\label{CosetTwist2}
S=
-\int d^2\sigma \left[A{\tilde L}^{a}_+{\tilde L}^{b}_-(V_k^T t^at^b V_k)+
B ({\tilde L}^{a(n)}_++R^{a(h)}_+)({\tilde L}^{b(n)}_-+R^{b(h)}_-)(V_k^T h^T t^at^b hV_k)\right].
\eea
Dualization of (\ref{CosetTwist2}) along $g$ directions gives
\bea\label{CosetTwist3}
S_{\text{dual}}=\int d^2\sigma \bigg[BR^a_+R^b_-(V^T_kh^Tt^at^bhV_k)+W_+^a(\mathcal{Y}^{-1})_{ab}W^b_-
\bigg].
\eea
To make the last expression more compact, we defined convenient quantities
\bea\label{CosetTwist4}
\mathcal{Y}^{ab}&=&A(V^T_kt^at^bV_k)+n^2B(V^T_kh^Tt^at^bhV_k)+{f_{ab}}^c u^c
\,,\nn
W^a_\pm&=&\d_\pm u^a\pm nB(V^T_kh^Tt^at^bhV_k)R^b_+\,.
\eea
The results (\ref{CosetTwist3})--(\ref{CosetTwist4}) is applicable to twisted versions (\ref{CosetTwist1}) of $S^k\times S^k$ for all values of $k$, and the final action has no continuous symmetries. To see this explicitly and to find a compact form of the dual metric and Kalb--Ramond field, we focus on $k=2$. In this case, various matrices appearing in (\ref{CosetTwist3})--(\ref{CosetTwist4}) are given by
\bea
V^T_2 h^Tt^at^b hV_2=-\frac{1}{2}\left[\begin{array}{ccc}
\sin^2\zeta&-\sin\zeta\cos\zeta&0\\
-\sin\zeta\cos\zeta&\cos^2\zeta&0\\
0&0&1
\end{array}\right],\
V_2^T t^b t^aV_2=-\frac{1}{2}\left[\begin{array}{ccc}
0&0&0\\
0&1&0\\
0&0&1
\end{array}\right]\,.\nonumber
\eea
This leads to a simple expression for the inverse of $(g+B)$ in the four--dimensional space spanned by 
$(\zeta,u^1,u^2,u^3)$:
\bea
(g+B)^{-1}=\begin{bmatrix}
-B^{-1}&0&0&\frac{n}{\sqrt{2}}\\
0&-\frac{1}{2}n^2B\sin{\zeta}^2&\sqrt{2}u_3+\frac{1}{4}n^2 B\sin{2\zeta}&-\sqrt{2}u_2 \\
0&-\sqrt{2}u_3+\frac{1}{4}n^2 B\sin{2\zeta}&-\frac{1}{4}(2A+n^2 B(1+\cos{2\zeta}))&\sqrt{2}u_1\\
-\frac{n}{\sqrt{2}}&\sqrt{2}u_2&-\sqrt{2}u_1&-\frac{A}{2}
\end{bmatrix}.
\eea
Similar expressions can be obtained for other values of $k$, but they involve the explicit form of structure constants for 
$SO(k+1)$, so we will not discuss them further.


\section{Discussion}
In this article we have used combinations of non--abelian T--dualities (NATD) and twists to generate new supergravity solutions from known geometries containing products of spheres. The abelian version of this procedure, TsT, has been successfully used in the past to generate geometries of black holes and gravity duals of various field theories, and the new non--abelian extension is expected to be fruitful as well. 

Let us briefly summarize our results. For geometries which contain at least one $S^3$ factor, we performed a sequence of NATDs  and explored an interplay between them in section \ref{SecS3}. In particular, we compared the duality procedures for groups and cosets, which have always been discussed separately in the literature, and found nice relations between them. Our findings are summarized in figure \ref{FigTriangle}. The solutions constructed in section \ref{SecS3} were obtained by dualizing the geometry (\ref{S3start}) along various directions, and each family had the same number of continuous parameters as the starting point (\ref{S3start}). To introduce additional parameters, one needs to mix several spheres, and the corresponding procedure was introduced in section \ref{SecSpecFlow}. In particular, we showed that such mixture, or twist, provides a natural generalization of the spectral flow operation from gravity duals of two--dimensional field theories to higher dimensional cases, and exploration of this new spectral flow on the field theory side is an interesting problem for future work. It is also intriguing to note that application of the twist puts the bubbling geometries describing giant gravitons in 10d in a form of fibration of $S^3\times R_t$ over a flat seven--dimensional base (\ref{SpecFlowBubble}). This fibration is reminiscent of one encountered in the two--charge D1--D5 geometries (\ref{D1D5twist}), and it suggests potential extensions of the geometries (\ref{SpecFlowBubble}) by allowing functions $Z_\pm$ in (\ref{ZpmDef})--(\ref{LaplaceZZ}) to depend on more than three coordinates on the base. In section \ref{SecSfTdGroup} we combined the geometrical twists with NATDs to introduce a new solution generating technique that generalizes TsT transformations, and we applied it to deform a large class of physically interesting solutions which have products of two three--spheres. Three--spheres are rather special since they can be viewed either as groups or as cosets, but for spaces containing $S^k\times S^k$ products for $k\ne 3$, only the second option is possible. Our general procedure of NATD--twist transformations summarized in section \ref{SecSfTdCoset} is applicable to such geometries as well. 

Several features of our NATD--twist procedure make it more challenging than the abelian TsT transformations. First, unlike the standard T--duality, its non--abelian version tends to destroy the symmetries of the original background making subsequent dualizations challenging. Second, twists (\ref{SpecFlowGH}) have nice geometrical interpretation only when parameter $n$ takes integer values, in contrast to the abelian case where one can shift by an arbitrary amount. Nevertheless, our final answers (\ref{gBmatrT1}) and (\ref{gBmatrT2a}), the results of the NATD--twist transformations parameterized by an integer $n$, can be formally viewed as supergravity solutions specified by a real $n$, putting the system on the same footing as in the abelian case. Application of subsequent dualities still remain challenging, and we hope to address it in future work.

In the abelian case, TsT transformations can be viewed as elements of a larger $O(d,d)$ group associated with reduction on a  torus and by embedding this group in the formalism of Double Field Theory. Recently NATD has been embedded in the DFT formalism as well \cite{DFT}, and it would be interesting to view our duality--twist transformations in this larger framework. We leave exploration of this embedding, as well as the relation between our construction and the algebraic classification of solution generating techniques developed in \cite{BDH21} for future work.


\section*{Acknowledgements}

This work was supported in part by the DOE grant DE-SC0015535.



\appendix
\section{Conventions}
In this appendix we collect some conventions used throughout the article. Specifically, we will focus on left invariant forms and their properties. Explicit expressions will be given for $SU(2)$ and $SO(k+1)/SO(k)$ cosets. We mostly follow the conventions of \cite{NATD101}.
\subsection{Group elements and left invariant forms}
\label{AppAsub1}
In this article we consider sigma models invariant under the left action of various groups. The relevant actions can be written in terms of the Maurer-Cartan left invariant forms
\begin{equation}
\begin{aligned}
L^i_\mu=-i\operatorname{Tr}(t^ig^{-1}\partial_\mu g),\quad L^i_\pm=L^i_\mu \partial_\pm X^\mu,\quad g\in G.
\end{aligned}    
\end{equation}
An example of an invariant action is
\begin{equation}
S=\int d^2\sigma E_{ij}L^i_+L^j_-,\quad E_{ij}=(g_{ij}+b_{ij})
\end{equation}
and some generalizations are discussed in the main body of this article. 
The generators of the group $t^i$ satisfy the Lie algebra with structure constants $f^{ij}{}_k$ defined by
\bea
[t^i,t^j]=f^{ij}{}_kt^k
\eea
To perform dualization along spheres $S^k$, we will be interested in Lie algebras for $SO(k+1)$ and their projections on $SO(k+1)/SO(k)$ cosets. In the $k=3$ case, we will also view the sphere as an $SU(2)$ group manifold.

For $G=SU(2)$, the generators are chosen to be $t^i=\sigma^i/\sqrt{2}$, where $\sigma^i$ are the Pauli matrices. This normalization implies that
\bea
\operatorname{Tr}(t^it^j)=\delta^{ij},\quad f_{ijk}=\sqrt{2}\epsilon_{ijk}\,.
\eea
While writing various actions, we also use a convenient matrix $D^{ij}=\operatorname{Tr}(t^igt^jg^{-1})$ defined in terms of a group element $g$, and in the normalization written above, this matrix is orthogonal: $D^TD=\mathbb{I}$.

To perform dualities along the $SO(N+1)/SO(N)$ cosets, we chose the generators of $SO(N+1)$ to be
\bea
J_{ij}\equiv t_{ij}-t_{ji}\,\quad \mbox{where}\quad  t_{ij(kl)}=\delta_{ik}\delta_{jl}\,,
\eea
with indices running over the values $1,2,\dots (N+1)$. The generators of the $SO(N)$ subgroup
are $J_{ab}$ with $a,b=1,2,\dots N$. For the $SO(4)/SO(3)$ coset, which is extensively discussed in section \ref{SecS3}, we use a special notation for the relevant generators: 
\begin{equation}
\begin{aligned}
 S_{a}&=J_{1,a+1}\quad\text{for~} a=1,2,3,\quad
S_{a+3}=J_{2,a+2}\quad\text{for~} a=1,2,\quad
S_{6}=J_{34}\,.
\end{aligned}
\end{equation}
We choose the $SO(3)$ subgroup to be spanned by $(A_1,A_2,A_3)=(S_6,S_5,S_4)$ and the coset generators to be spanned by $(B_1,B_2,B_3)=(S_1,S_2,S_3)$. This leads to simple commutation relations
\begin{equation}
[A_a,A_b]=\epsilon_{abc}A_c\quad [B_a,B_b]=\epsilon_{abc}A_c \quad
[A_a,B_b]=\epsilon_{abc}B_c,\quad \text{i.e.~}[S_i,S_j]=\tensor{f}{_{ij}^k}S_k
\end{equation}
For completeness we summarize the explicit form of matrices $S_k$:
\bea
&&B_1=  S_1=\begin{bmatrix}
0&0&0&-1\\
0&0&0&0\\
0&0&0&0\\
1&0&0&0\\
\end{bmatrix}\quad B_2= S_2=\begin{bmatrix}
0&0&0&0\\
0&0&0&-1\\
0&0&0&0\\
0&1&0&0
\\
\end{bmatrix}\quad B_3=S_3=\begin{bmatrix}
0&0&0&0\\
0&0&0&0\\
0&0&0&-1\\
0&0&1&0\\
\end{bmatrix}\nn
\\
&&A_3=S_4=\begin{bmatrix}
0&-1&0&0\\
1&0&0&0\\
0&0&0&0\\
0&0&0&0\\
\end{bmatrix}\quad A_2=S_5=\begin{bmatrix}
0&0&1&0\\
0&0&0&0\\
-1&0&0&0\\
0&0&0&0\\
\end{bmatrix}\quad A_1=S_6=\begin{bmatrix}
0&0&0&0\\
0&0&-1&0\\
0&1&0&0\\
0&0&0&0\\
\end{bmatrix}\nonumber
\eea
Similarly for the case of $S^2$, which is explored in section \ref{SecSubSU2dbl}, the relevant generators $S_k$ and $A_1$ are
\begin{equation}
\begin{aligned}
& A_1=S_3=\begin{bmatrix}
0&-1&0\\
1&0&0\\
0&0&0\\
\end{bmatrix}\,,\quad S_2=\begin{bmatrix}
0&0&1\\
0&0&0\\
-1&0&0\\
\end{bmatrix}\,,\quad S_1=\begin{bmatrix}
0&0&0\\
0&0&-1\\
0&1&0\\
\end{bmatrix}\,,    
\end{aligned}    
\end{equation}
and they obey the $SO(3)$ algebra $[S_i,S_j]=\epsilon_{ijk}S_k$.

\subsection{Twisted versions of the left invariant forms}
\label{AppSubLn}

While studying twists of spheres in section \ref{SecSpecFlow}, we introduced convenient objects 
${\tilde L}^{{a}(n)}$ which generalize the standard left invariant forms ${\tilde L}^{{a}}$:
\bea\label{LformsOneApp}
{\tilde L}^{{a}}=-i\mbox{Tr}(t^{{a}}g^{-1}dg),\quad
{\tilde L}^{{a}(n)}=-i\mbox{Tr}(t^{{a}}g^{-n}d(g^n))
\eea
In this appendix we will derives some properties of ${\tilde L}^{{a}}$ which are used throughout the article. We begin with observing that ${\tilde L}^{{a}(1)}={\tilde L}^{{a}}$ and applying induction\footnote{Since extensions (\ref{LformsOneApp}) can be formulated both for ${\tilde L}^{{a}}$ and 
${L}^{{k}}=-i\mbox{Tr}(t^{{k}}h^{-1}dh)$, in this appendix we use neutral notation $p$ for the group element. By setting $p=g$ we would get ${\tilde L}^{{a}(n)}$, and by setting $p=h$ we would get ${L}^{{k}}$.}:
\begin{equation}\label{induction left invariant C}
L^{(n)}_i=-i\operatorname{Tr}(t_ip^{-n}dp^n)=L^{(n-1)}_i-i\operatorname{Tr}(t_ip^{-n}dpp^{n-1})    
\end{equation}
Observing that the matrix $pt_ip^{-1}$ is Hermitian and traceless, we conclude that $pt_i   p^{-1}$ gives some rotation of the basis $t_j$:
\begin{equation}
pt_i   p^{-1} =R_{ij}t_j,\quad R_{ij}=\operatorname{Tr}(pt_ip^{-1}t_j)=D_{ji}
\end{equation}
Note that the matrix $R$ also appears in the $(g+B)^{-1}$ matrix (\ref{gBmatrT2a}). Furthermore, since $RR^T=\mathbb{I}$, 
\begin{equation}
p^{-1}t_ip=(R^{-1})_{ij}t_j    
\end{equation}
Using induction, we compute
\bea
-i\mbox{Tr}[t_ip^{-n}dpp^{n-1}]&=&R_{ij}-i\mbox{Tr}[t_jp^{-n+1}dpp^{n-2}]=\ldots= R^{n-1}_{ij}-i\mbox{Tr}[t_jp^{-1}dp]\nn
&=&(R^{n-1})_{ij}L_j    
\eea
Substitution of this result into (\ref{induction left invariant C}) gives a compact expression for $L^{(n)}_i$:
\bea
L^{(n)}_i=L^{(n-1)}_{i}+(R^{n-1})_{ij}L_j=\ldots=\big[\mathbb{I}+R+\ldots+R^{n-1}\big]_{ij}L_j=\Bigg[\frac{\mathbb{I}-R^n}{\mathbb{I}-R}\Bigg]_{ij}L_j 
\eea
This expression applies to all the positive integers $n$, and extension this to negative values of $n$ comes from similar arguments:
\begin{equation}
\begin{aligned}
&L_i^{-1}=-i\mbox{Tr}[t_ipdp^{-1}]=i\operatorname{Tr}[t_idpp^{-1}]=-(R^{-1})_{ij}L_j\\
&L^{(-n)}_i=-i\mbox{Tr}[t^ip^ndp^{-n}]=i\mbox{Tr}[t_idp^np^{-n}]=-(R^{-n})_{ij}L^{(n)}_j=\Bigg[\frac{\mathbb{I}-R^{-n}}{\mathbb{I}-R}\Bigg]_{ij}L_j
\end{aligned}
\end{equation}
Here we used the following relations
\bea
p^{-n}t_ip^n=(R^{-1})_{ij}p^{-n+1}t_j p^{n-1}=\ldots=(R^{-n})_{ij} t_j\,.\nonumber
\eea
Therefore, for all integer values of $k$, we have,
\begin{equation}
L^{(k)}_i=\Bigg[\frac{\mathbb{I}-R^k}{\mathbb{I}-R}\Bigg]_{ij}L_j   
\end{equation}
These relations are used in sections \ref{SecSpecFlow} and \ref{SecSfTdGroup}.

\section{KK reduction in the metric and in the action}
\label{B}

In section \ref{SecSfTdGroup} we found that some results of NATD can be written compact forms 
by combining the NS--NS fields into a matrix $(g+B)^{-1}$. Some examples are given by (\ref{gBmatrT1}) and (\ref{OptBdualNS}). On the other hand, the expressions for the metric, e.g.,
(\ref{SolnTdualTrivD}) and (\ref{SolnTlNonTrivD}), are rather complicated. In this short appendix we summarize the general relations between Kaluza--Klein reduction in the metric and in the sigma model action, which clarify the map between the metric components and $(g+B)^{-1}$.

Let us consider a metric and a B field written as fibrations over the space spanned by coordinates $v_k$:
\bea\label{KKmetr}
ds^2&=&g_{ij}(dv^i+A^i)(dv^j+A^j)+g_{ab}dz^adz^b\nn
B&=&b_{ij}(dv^i+A^i)\wedge (dv^j+A^j)+2dv^i\wedge B_i+b_{ab}dz^a\wedge dz^b
\eea
Solutions of this type are discussed in section \ref{SecSfTdGroup}, where $v_k$ represent the coordinates after non--abelian T duality. Combining the expressions above into an action, we find
\bea
S=\int d^2\sigma\left[(g_{ij}+b_{ij})(\d_+v^i+A_+^i)(\d_-v^j+A_-^j)+\d_+ v^iB_{i-}-\d_- v^iB_{i+}+{\tilde q}_{ab}\d_+z^a\d_-z^b\right]\nonumber
\eea
Completing the square in the cross terms, we arrive at a compact form of the action
\bea\label{KKaction}
S=\int d^2\sigma\left[(g_{ij}+b_{ij})(\d_+v^i+{\cal A}_+^i)(\d_-v^j+{\cal A}_-^j)+q_{ab}\d_+z^a\d_-z^b\right]
\eea
and the relations between the ingredients of (\ref{KKmetr}) and (\ref{KKaction})
\bea
{\cal A}_+^i=A_+^i-G^{ij}B_{j+},\quad {\cal A}_-^i=A_-^i+G^{ij}B_{j-},\quad G_{ij}=g_{ij}+b_{ij}\,.
\eea
In section \ref{SecSfTdGroup} we will also need the inverse relation:
\bea
A^i_\mu=\frac{1}{2}g^{ik}\left[G_{kj}{\cal A}_{-\mu}^j+G_{jk}{\cal A}_{+\mu}^j\right]\,.
\eea

\section{Ramond--Ramond fields for the dual backgrounds}
\label{SecAppRR}

In section \ref{SecSfTdGroup} we applied various non--abelian dualities to the system (\ref{SpecFlowBubble}). In this appendix we will present some technical details of the relevant procedures and derive the expressions for the Ramond--Ramond fields. In section \ref{SecSubAppRRrev} we review the general procedure for dualizing RR fluxes, and in the remaining part of this appendix this construction will be applies to geometries with twisted product of three--spheres.

\subsection{Review of the general procedure for dualizing along $SU(2)$}
\label{SecSubAppRRrev}
In this subsection we review the general procedure for dualizing Ramond-Ramond fields on backgrounds with an $SU(2)$ isometry following \cite{NATD101}. 

Consider a sigma model with a Lagrangian density 
\begin{equation}\label{SigmaModL}
\begin{aligned}
\mathcal{L}_0&=Q_{\mu\nu}\d_+X^\mu \d_-X^\nu+Q_{\mu i}\d_+X^\mu L_-^i+
Q_{i\mu }L_+^i \d_-X^\mu+E_{ij}L^i_+L^j_-\,,
\end{aligned}
\end{equation}
where
\begin{equation}
\begin{aligned}
Q_{\mu\nu}=G_{\mu\nu}+B_{\mu\nu},\quad Q_{\mu i}=G_{\mu i}+B_{\mu i },\quad Q_{i\mu}=G_{i\mu}+B_{i\mu},\quad E_{ij}=g_{ij}+b_{ij}
\end{aligned}    
\end{equation}
and $L_-^i$ are left invariant forms of $SU(2)$. The RR fluxes supporting the geometry (\ref{SigmaModL}) will be specified below. To perform an NATD along $SU(2)$, one begins with decomposing the frames for the metric (\ref{SigmaModL}) as
\begin{equation}\label{StartFrameApp}
e^a=\tensor{\kappa}{^a_i}L^i+\tensor{\lambda}{^a_\mu}dx^\mu,\quad e^A=\tensor{e}{^A_\mu}dx^\mu\,,
\end{equation}
where $i=1,2,3$, and index $A$ runs over the remaining seven directions. Matrices 
$(\tensor{\kappa}{^a_i},\tensor{\lambda}{^a_\mu})$ are determined by solving quadratic relations 
\begin{equation}
\tensor{\kappa}{^a_i}\tensor{\kappa}{^a_j}=g_{ij},\quad \tensor{\kappa}{^a_i}\tensor{\lambda}{^a_\mu}=G_{i\mu},\quad \tensor{\lambda}{^a_\mu}\tensor{\lambda}{^a_\nu}=K_{\mu\nu},\quad \eta_{AB}\tensor{e}{^A_{\mu}}\tensor{e}{^B_{\nu}}=G_{\mu\nu}-K_{\mu\nu}
\end{equation}
The transformation matrix $\Omega$ defined by (\ref{Omdefn}) can be written as
\begin{equation}\label{OmegaApp}
\begin{aligned}
\Omega^{-1}=(A_0\Gamma^1\Gamma^2\Gamma^3+A_a\Gamma^a)\Gamma_{11},\quad A_0=\frac{1}{\sqrt{1+\zeta^2}},\quad A_a=\frac{\zeta^a}{\sqrt{1+\zeta^2}}    
\end{aligned}
\end{equation}
where
\bea
\zeta^a=\kappa^a{}_iz^i,\quad z^i=\frac{(\frac{1}{2}\eps_{ijk}b_{jk}+\sqrt{2}v_i)}{\sqrt{\operatorname{det}g}}
\eea
Let us assume that the original geometry (\ref{SigmaModL}) is supported by Ramond--Ramond fluxes which have the form
\begin{equation}\label{FasGapp}
\begin{aligned}
F_p&=G^{(0)}_p+G^a_{p-1}\wedge e^a+\frac{1}{2}G^{ab}_{p-2}\wedge e^a\wedge e^b+G^{(3)}_{p-3}\wedge e^1\wedge e^2\wedge e^3 \\
\slashed{F}_p&=\slashed{G}^{(0)}_p\mathbb{I}+\slashed{G}^a_{p-1} \Gamma^a+\frac{1}{2}\slashed{G}^{ab}_{p-2}\Gamma^{ab}+\slashed{G}^{(3)}_{p-3}\Gamma^{123}
\end{aligned}
\end{equation}
Then application of the general procedure (\ref{RRproj})--(\ref{RR trans defn}) leads to the dual fluxes ${\hat F}$:
\begin{equation}
\slashed{F}_p\Omega^{-1}=\hat{\slashed{F}}_{p-3}+\hat{\slashed{F}}_{p-1}+\hat{\slashed{F}}_{p+1}+\hat{\slashed{F}}_{p+3}
\end{equation}
Substitution of (\ref{OmegaApp}) and (\ref{FasGapp}) into these expressions leads to the final result for the dual fluxes
\begin{equation}
\begin{aligned}
\hat{\slashed{F}}_{p-3}&=-A_0\slashed{G}^{(3)}_{p-3},\\
\hat{\slashed{F}}_{p-1}&=A_a\slashed{G}^a_{p-1}-\frac{A_0}{2}\slashed{G}^{ab}_{p-2}\epsilon^{abc}\Gamma^c-{A_a}\slashed{G}^{ab}_{p-2}\Gamma^b+\frac{A_a}{2}\slashed{G}^{(3)}_{p-3}\epsilon^{abc}\Gamma^{bc},\\
\hat{\slashed{F}}_{p+1}&=A_a\slashed{G}^{(0)}_{p}\Gamma^a+\frac{A_0}{2}\slashed{G}^{a}_{p-1}\epsilon^{abc}\Gamma^{bc}-{A_a}\slashed{G}^{b}_{p-1}\Gamma^{ab}+\frac{A_a}{2}\slashed{G}^{(bc)}_{p-2}\epsilon^{abc}\Gamma^{123},\\
\tilde{\slashed{F}}_{p+3}&=A_0\slashed{G}^{(0)}_{p}\Gamma^{123}.
\end{aligned}
\end{equation}
Decomposing thee dual RR fields in a fashion similar to (\ref{FasGapp}),
\begin{equation}\label{hat dual RR}
\hat{F}_p=\hat{G}^{(0)}_p  +\hat{G}^a_{p-1}\wedge \hat{e}^a+\frac{1}{2}\hat{G}^{ab}_{p-2}\wedge \hat{e}^a\wedge \hat{e}^b+\hat{G}^{(3)}_{p-3}\wedge \hat{e}^1\wedge \hat{e}^2\wedge \hat{e}^3  
\end{equation}
we deduce the expressions for various ingredients:
\begin{equation}\label{Ghat}
\begin{aligned}
\hat{G}^{(0)}_p&=e^{\phi-\hat{\phi}}\bigg(-A_0G_p^{(3)}+A_aG^a_p\bigg),\\
\hat{G}^{a}_{p-1}&=e^{\phi-\hat{\phi}}\bigg(-\frac{A_0}{2}\epsilon^{abc}G^{bc}_{p-1}+A_bG^{ab}_{p-1}+A_aG^{(0)}_{p-1}\bigg),\\
\hat{G}^{ab}_{p-2}&=e^{\phi-\hat{\phi}}\bigg[\epsilon^{abc}\big(A_cG^{(3)}_{p-2}+A_0G^c_{p-2}\big)-\big(A_aG^{b}_{p-2}-A_bG^{a}_{p-2}\big)\bigg],\\
\hat{G}^{(3)}_{p-3}&=e^{\phi-\hat{\phi}}\bigg(\frac{A_a}{2}\epsilon^{abc}G^{bc}_{p-3}+A_0G^{(0)}_{p-3}\bigg).
\end{aligned}    
\end{equation}
Recall that the frame after duality are given by (\ref{DualFrames}). In the remaining part of this appendix we will use equations (\ref{hat dual RR}) and (\ref{Ghat}) to dualize RR fluxes on twisted versions (\ref{SpecFlowBubble}) of bubbling geometries (\ref{BG metric}).

\subsection{Option A}
\label{AppRRopA}

In section \ref{SecSubOptA} we started with action (\ref{ActionSF}) and applied the non--abelian duality along directions $h$ associated with one of the two $SU(2)$s. Let us assume that the initial metric is supported by the fluxes originating from the twisted bubbling geometries (\ref{SpecFlowBubble}):
\bea
F_5={G}_2\wedge (L+{\tilde L}^{(n)}D)^1\wedge (L+{\tilde L}^{(n)}D)^2\wedge (L+{\tilde L}^{(n)}D)^3+
\tilde{G}_2\wedge {\tilde L}^{\tilde{1}}\wedge {\tilde L}^{\tilde{2}}\wedge {\tilde L}^{\tilde{3}}
\eea
To perform the duality, we begin with rewriting this expression in terms of the appropriate frames:
\bea
F_5&=&
\frac{G_2}{E\sqrt{E}}\wedge e^1\wedge e^2\wedge e^3+
\tilde{G}_2\wedge {\tilde L}^{\tilde{1}}\wedge {\tilde L}^{\tilde{2}}\wedge {\tilde L}^{\tilde{3}}\nn
e^a&=&\sqrt{E}(L_++\tensor{D}{}\tilde{L}_+^{(n)})^a
\eea
Then application of the procedure outlined in section \ref{SecSubAppRRrev} gives the Ramond--Ramond for the dual geometry (\ref{SolnTdualTrivD}):
\bea\label{StrengthOptAstart}
\hat{F_2}=-\frac{G_2}{8},\quad
\hat{F}_4=\frac{1}{2}\hat{G}^{ab}_2\wedge \hat{e}^a\wedge \hat{e}^b,\quad
\hat{F}_6=\hat{G}^a_{5}\wedge \hat{e}^a,\quad
\hat{F}_8=\hat{G}^{(3)}_5\wedge\hat{e}^1\wedge \hat{e}^2\wedge \hat{e}^3  
\eea
with
\bea
\hat{G}^{ab}_2=\frac{\sqrt{2}}{E}\frac{G_2}{8}\epsilon_{abc}v^c,\quad
\hat{G}^a_5=\frac{\sqrt{2E}}{8}v^a(\tilde{G_2})(\tilde{L}^{1}\tilde{L}^{2}\tilde{L}^{3}),\quad
\hat{G}^{(3)}_5=\frac{{E\sqrt{E}}}{8}(\tilde{G}_2)(\tilde{L}^{1}\tilde{L}^{2}\tilde{L}^{3})
\eea
The dual frames in (\ref{StrengthOptAstart}) are given by
\bea
\hat{e}^a&=&\frac{1}{\sqrt{E}(E^2+2r^2)}\bigg[E[\tilde{L}^{a(n)}(2r^2)-2v^a(\tilde{L}^{i(n)}v_i)-\sqrt{2}\epsilon_{aij}dv^iv^j]-2v^a(v^idv_i)\nn
&&-E^2[dv^a+\sqrt{2}\epsilon_{aij}\tilde{L}^{i(n)}v^j]\bigg].
\eea
To proceed we will also need the expressions for several products:
\bea
\hat{e}^a\wedge \hat{e}^b&=&\frac{1}{(E^2+2r^2)}\bigg[(E)dv^adv^b+\sqrt{2}\epsilon_{abc}(dv_iv^i)dv^c+2\epsilon_{abc}v^c\bigg(\frac{E}{2}\epsilon_{kli}\tilde{L}^{k(n)}\tilde{L}^{l(n)}v^i\bigg)\nn
&&+\sqrt{2}E\epsilon_{aij}dv^bv^i\tilde{L}^{j(n)}-\sqrt{2}E\epsilon_{bij}dv^av^i\tilde{L}^{j(n)}+2(dv_xv^x)(\tilde{L}^{a(n)}v^b-\tilde{L}^{b(n)}v^a)\bigg]\nn
\hat{e}^1\wedge \hat{e}^2\wedge \hat{e}^3&=&\frac{-\sqrt{E}}{E^2+2r^2}\big[dv_1dv_2dv_3+\sqrt{2}(dv_av^a)(\tilde{L}^{b(n)}dv_b)+(dv_av^a)(\epsilon_{ijk}\tilde{L}^{i(n)}\tilde{L}^{j(n)}v^k)\big]    \nn
\epsilon_{abc}v^c\hat{e}^a\wedge \hat{e}^b&=&\frac{\sqrt{2}E}{(E^2+2r^2)}\big[\frac{1}{2}\epsilon_{abc}[\sqrt{2}dv^adv^bv^c+{\sqrt{2}}\tilde{L}^{a(n)}\tilde{L}^{b(n)}v^c(2r^2)]-(dv_a\tilde{L}^{a(n)})(2r^2)\nn
&&+2(dv_av^a)(\tilde{L}^{b(n)}v_b)\big]\nonumber
\eea
Substitution of these products into (\ref{StrengthOptAstart}) leads to rather complicated expressions for field strengths,
\bea\label{StrengthOptA}
\hat{F}_2&=&-\frac{G_2}{8},\quad
\hat{F}_6=-\frac{\sqrt{2}\tilde{G}_2}{8}(\tilde{L}^{1}\tilde{L}^{2}\tilde{L}^{3})(v_idv^i),
\nn
\hat{F}_4&=&\frac{G_2}{8(E^2+2r^2)}\Big[\frac{1}{\sqrt{2}}\epsilon_{abc}[dv^adv^bv^c+\tilde{L}^{a(n)}\tilde{L}^{b(n)}v^c(2r^2)]\\
&&
-(dv_a\tilde{L}^{a(n)})(2r^2)
+2(dv_av^a)(\tilde{L}^{b(n)}v_b)\Big]\nn
\hat{F}_8&=&-\frac{E^2}{8(E^2+2r^2)}(\tilde{G}_2)(\t{L}^1\t{L}^2\t{L}^3)\big[dv_1dv_2dv_3\big]\nonumber
\eea
but gauge potentials are remarkably simple. To extract them, we recall the general relations
\bea\label{ChernF246}
&&d\t{C}_1=\hat{F}_2,\quad d\t{C}_3=\hat{F}_4-B\wedge \hat{F}_2,\quad
d\t{C}_5=\hat{F}_6-B\wedge \hat{F}_4+\frac{1}{2}B^2\wedge\hat{F}_2,\nn
&&d\t{C}_7=\hat{F}_8-B\wedge \hat{F}_6+\frac{1}{2}B^2\wedge \hat{F}_4-\frac{1}{6}B^3\wedge\hat{F}_2
\eea
Rewriting the Kalb--Ramond field (\ref{SolnTdualTrivD}) in terms of frames,
\bea\label{BfieldOptA}
B=-\frac{\sqrt{2}}{2E}\epsilon_{abc}\hat{e}^a\wedge\hat{e}^bv^c+\frac{\sqrt{2}}{2}\epsilon_{abc}\t{L}^{a(n)}\t{L}^{b(n)}v^c+(\t{L}^{a(n)}dv_a),
\eea
substituting the result into (\ref{ChernF246}), and simplifying the final expressions, we find exterior derivatives of the RR potentials:
\bea\label{PotOptA}
&&d\t{C}_1=-\frac{G_2}{8},\quad d\t{C}_3=d\left[\frac{G_2}{8}\t{L}^{a(n)}v_a\right]    \nn
&&d\t{C}_5=
\frac{G_2}{16}d\left[v_a\t{L}^{a(n)}\wedge (\t{L}^{b(n)}dv_b)\right]-\frac{\sqrt{2}\tilde{G}_2}{16}(\tilde{L}^{1}\tilde{L}^{2}\tilde{L}^{3})d(v_av_a)\\
&&d\t{C}_7=\left[-\frac{\t{G}_2}{8}(\t{L}^1\t{L}^2\t{L}^3)-\frac{G_2}{8}(\t{L}^{1(n)}\t{L}^{3(n)}\t{L}^{3(n)})\right](dv_1dv_2dv_3)  
\nonumber
\eea

\subsection{Option B}
\label{AppRRoptB}
In section \ref{SecSubOptB} we dualized the geometry (\ref{ActionSF}) along $g$ direction and derived the expressions for the NS--NS fields. In this appendix we present the derivation of the Ramond--Ramond fluxes supporting the dual geometry.

We begin with recalling that, while the $SU(2)$ symmetry associated with $h$ directions is present in the twisted bubbling geometries (\ref{SpecFlowBubble}) for an arbitrary 
$n$, the symmetry associated with rotation of $g$ is present only for $n=0,\pm1$ (see table \ref{SymGroupTable}). In the $n=0$ case, a $Z_2$ symmetry between two three--dimensional spheres exchanges $g$ and $h$ directions, so the fluxes for option B can be easily extracted from our earlier discussion of option A. The derivations of fluxes for $n=1$ and $n=-1$ follow the same logic, so here we will focus only on the former case.
 
We begin with recalling the frames before the duality and expressing the solution for the integrated gauge field in terms of them:
\bea
e^a&=&\frac{1}{\sqrt{H}}(H\t{L}^a+EL^a),\quad Y=\sqrt{H^2+2r^2},\quad Z=\sqrt{H}Y\\
A^a&=&\frac{\sqrt{2}s^a}{Y},\quad A_0=\frac{H}{Y},\quad e^{\phi-\hat{\phi}}=Z
\eea
Rewriting the five--form flux (\ref{SpecFlowBubble}) in terms of these frames as
\bea
F_5=G^{(0)}_5+G^a_{4}\wedge e^a+\frac{1}{2}G^{ab}_3\wedge e^a\wedge e^b+G^{(3)}_{2}\wedge e^1\wedge e^2\wedge e^3
\eea
we extract various ingredients:
\bea
&&G^{(0)}_5=\frac{(\t{G_2}(-E)^3+F^3G_2)}{H^3}L^1L^2L^3,\quad
G^a_4=\frac{\sqrt{H}}{H^3}(\t{G}_2(E)^2+G_2F^2)\frac{1}{2}\epsilon_{abc}L^bL^c\nn
&&G^{ab}_3=\frac{H}{H^3}(\t{G}_2(-E)+FG_2)\epsilon_{eab}L^e,\quad
G^{(3)}_2=\frac{(\t{G}_2+G_2)H^{3/2}}{H^3}
\eea
This leads to the expressions for the RR field strengths after duality
\bea\label{StenOptB}
\hat{F}_2&=&e^{\phi-\hat{\phi}}(-A_0G^{(3)}_2)=-\frac{1}{8}(\t{G}_2+G_2)\nn
\hat{F}_4&=&\hat{G}^{(0)}_4+\hat{G}^a_{3}\wedge \hat{e}^a+\frac{1}{2}\hat{G}^{ab}_2\wedge \hat{e}^a\wedge \hat{e}^b\\
\hat{F}_6&=&\hat{G}^a_{5}\wedge \hat{e}^a+\frac{1}{2}\hat{G}^{ab}_4\wedge \hat{e}^a\wedge \hat{e}^b+\hat{G}^{(3)}_{3}\wedge \hat{e}^1\wedge \hat{e}^2\wedge \hat{e}^3\nonumber\\
\hat{F}_8&=&\hat{G}^{(3)}_5\wedge\hat{e}^1\wedge \hat{e}^2\wedge \hat{e}^3\nonumber
\eea
where the dual frames are given by
\bea
\hat{e}^a&=&\frac{1}{\sqrt{H}Y^2}\bigg[-2s^a(s_ids^i)-2 Es^a(s_iL^i)-\sqrt{2}H\epsilon_{axy}ds^xs^y\nn
&&+[E(2s_is^i)]{L}^a-H^2ds^a-\sqrt{2}EH\epsilon_{aij}L^is^j\bigg]
\eea
The duality rules reviewed in section \ref{SecReview} give 
\bea\label{StenOptBe2}
&&{\hat G}^{(0)}_4=\frac{\sqrt{2}}{16H^2}(\t{G}_2E^2+G_2F^2)\epsilon_{abc}L^aL^bs^c,\quad
{\hat G}^{ab}_2=\frac{\sqrt{2}}{8H}{(G_2+\tilde{G}_2)}\epsilon_{abc}s_c\,,\nn
&&{\hat G}^{a}_3=\frac{-\t{G}_2E+FG_2}{8\sqrt{H}}\bigg[-L^a+\frac{\sqrt{2}}{H}\epsilon_{iab}L^is^b\bigg]\,,\quad
\hat{G}^a_5=\frac{-\t{G_2}E^3+F^{3}G_2}{4\sqrt{2}H^{5/2}}\bigg[L^1L^2L^3s^a\bigg]\,,\nn
&&\hat{G}^{ab}_4=\frac{1}{16H^2}(\t{G}_2E^2+F^2G_2)\bigg[L^iL^j(2H\delta_i^a\delta_j^b-\sqrt{2}s^a\epsilon_{ijb}+\sqrt{2}s^b\epsilon_{ija})\bigg]\,,\\
&&\hat{G}^{(3)}_3=\frac{-\t{G}_2 E+FG_2}{4\sqrt{2}H^{3/2}}s_aL^a,\quad
\hat{G}^{(3)}_5=\frac{-\t{G_2}E^3+F^3G_2}{{H}Y^2}(L^1L^2L^3)(ds_1ds_2ds_3)\,.
\nonumber
\eea
Substituting these ingredients into the relations (\ref{ChernF246}) and using the expression (\ref{SolnTlNonTrivD}) for the Kalb--Ramond field,
\bea
B&=&\frac{1}{Y^2}\bigg[\frac{\sqrt{2}E^2}{(2)}\epsilon_{abc}L^aL^bs^c-\frac{\sqrt{2}}{(2)}\epsilon_{abc}ds^ads^bs^c+{EH}(L_ids^i)+\frac{2E}{H}(L_is^i)(s^ads^a)  \bigg]  \nn
&=&-\frac{\sqrt{2}}{2H}\epsilon_{abc}\hat{e}^a\wedge \hat{e}^bs^c+\frac{\sqrt{2}}{2H^2}(E^2)\epsilon_{aij}L^aL^is^j+\frac{E}{H}(L^ads_a),
\eea
we find the exterior derivatives of the RR potentials:
\bea\label{StenOptBe3}
&&d\t{C}_1=-\frac{1}{8}(\t{G}_2+G_2),\quad
d\t{C}_3=-d\left[\frac{G_2}{8}(L^as_a)\right],\nn
&&d\t{C}_5=d\left[\frac{G_2}{2\times 8}(L^as_a)(L^ids_i)\right],\quad
d\t{C}_7=-\frac{G_2}{8}(L_1L_2L_3)(ds_1ds_2ds_3)
\eea

\subsection{Option C}
\label{AppRRoptC}

Option C was introduced in section \ref{SecSubOptC} as a sequence of dualities along $g$ and $h$ directions. Although the final geometry is the same regardless of the order of dualization, the 
detailed construction of the RR fields is sensitive to the order, so it is instructive to discuss both options. 

We begin with dualizing the result of the option A along $g$ direction. To construct the dual RR fluxes, we need to construct the frames for the starting geometry, i.e., for the outcome of the option A. Starting with the metric and the B field,
\bea
(F+E-E\hat{M}^{-1}E)_{ij}=g_{ij}+b_{ij}\,,
\eea
and writing them as
\bea
g_{ij}=\kappa^a{}_i\kappa^a{}_j,\quad b_{ij}=\epsilon_{ijk}b_k\,,
\eea
we extract the matrix involved in dualization of the RR fields (see (\ref{OmegaApp})):
\begin{equation}
\Omega_{t_2}^{-1}=(A_0\Gamma^{123}+A_{1a}\Gamma^a)\Gamma_{11},\quad A_0=\frac{1}{\sqrt{1+\zeta^2}},\quad A_1=\frac{\zeta^a}{\sqrt{1+\zeta^2}}    
\end{equation}
Here parameters $\zeta^a$ are defined in terms of $(\kappa_{ij},b_k)$ and the dual coordinates $s_i$:
\bea
\zeta^a=\kappa^a{}_iz^i\,,\quad z^i=\frac{(b_i+\sqrt{2}s_i)}{\sqrt{\operatorname{det}g}}\,.
\eea
Using the explicit expressions
\bea
&&\kappa^a{}_i=\frac{1}{\gamma}\begin{pmatrix}
\alpha+\beta v_{\t{1}}^2& \beta v_{\t{1}}v_{\t{2}}&\beta v_{\t{1}}v_{\t{3}}\\
\beta v_{\t{1}}v_{\t{2}}&\alpha +\beta v_{\t{2}}^2&\beta v_{\t{2}}v_{\t{3}}\\
\beta v_{\t{1}}v_{\t{3}}&\beta v_{\t{2}}v_{\t{3}}&\alpha +\beta v_{\t{3}}^2
\end{pmatrix},\quad
b_k=\frac{E^2\sqrt{2}v_{\t{k}}}{E^2+2r^2}
 \nn
&&\gamma=\sqrt{E^2+2r^2},\quad
\beta=-\frac{2}{2r^2}\sqrt{F(E^2+2r^2)}+  \sqrt{E^2F+2(E+F)^2r^2}\\
&&\alpha=\sqrt{E^2F+2(E+F)^2r^2}\nonumber
\eea
as well the determinant of the metric,
\bea
\sqrt{\operatorname{det}g}=\sqrt{F}\bigg[F+\frac{2E r^2}{E^2+2r^2}\bigg],
\eea
one can write $\Omega_{t_2}$ in terms of $(E,F,r,v_k)$, but the final result is not very illuminating. 
The frames after duality are given by
\begin{equation}
\bar{e}=-\kappa W^{-T}(ds-\hat{M}^{-1}Edv)+\lambda dv    
\end{equation}
where $\kappa$ is given above, and other ingredients are
\begin{equation}
\hat{M}=E+f_v,\quad \lambda=-\kappa^{-1}\hat{M}^{-1}E,\quad W=[H-E\hat{M}^{-1}E+f_s],\quad {f_{s(v)}}_{ab}={s(v)}_kf^k_{ab}   
\end{equation}
The RR fluxes constructed in option A (\ref{StrengthOptA}) are rewritten using the frames (\ref{StartFrameApp}), to get the contributions relevant in performing the second duality along $\tilde{L}^a$ directions, such that
\bea
F_p=G^{0}_p+G^a_{p-1}\wedge e^a+\frac{1}{2}G^{ab}_{p-2}\wedge e^a\wedge e^b+G^{(3)}_{p-3}\wedge e^1\wedge e^2\wedge e^3    
\eea
For the option A RR fluxes we obtain
\begin{equation}\label{flux Gs A-1}
\begin{aligned}
G^{(3)}_3&=\frac{\sqrt{2}\tilde{G}_2}{8}(\operatorname{Det}A)(v_idv^i),\quad G^{ab}_4=\frac{2\sqrt{2}\tilde{G}_2}{8}(v_ldv^l)\epsilon_{ijk}(\tensor{A}{_a^i}\tensor{A}{_b^j}\tensor{B}{_c^k})(dv^c)\\
G^{(3)}_5&=\frac{E^2}{8(E^2+2r^2)}(\tilde{G}_2)(\operatorname{Det}A)(dv^1dv^2dv^3)\\
G^{(0)}_2&=\frac{-G_2}{8},\quad G^c_5=\frac{-\sqrt{2}\tilde{G}_2}{8}\epsilon_{ijk}(\tensor{B}{_a^i}\tensor{B}{_b^j}\tensor{A}{_c^k})(v_ldv^l)(dv^adv^b)\\
G^{(0)}_4&=\frac{G_2}{8(E^2+2r^2)}\bigg[\frac{{1}}{\sqrt{2}}\epsilon_{abc}\big[dv^adv^bv^c+(\tensor{B}{_i^a}dv^i)(\tensor{B}{_j^b}dv^j)v^c(2r^2)\big]\\
&+\tensor{B}{_i^a}(dv_adv^i)(2r^2)-2\tensor{B}{_i^b}(dv^av_a)(dv^iv_b)\bigg]\\
G^j_3&=\frac{-G_2}{8(E^2+2r^2)}\bigg[\sqrt{2}\epsilon_{abc}(\tensor{B}{_i^a}dv^i)\tensor{A}{_j^b}v^c(2r^2)+(2r^2)dv_a\tensor{A}{_j^a}-2(dv^av_a)(\tensor{A}{_j^b}v_b)\bigg]\\
G^{ij}_2&=\frac{\sqrt{2}G_2}{8(E^2+2r^2)}\epsilon_{abc}\tensor{A}{_i^a}\tensor{A}{_j^b}v^c(2r^2)\\
\end{aligned}
\end{equation}
Here we have redefined $A\equiv\kappa^{-1},~B\equiv\kappa^{-1}\lambda$. These contributions result in the dual RR fluxes when substituted in (\ref{Ghat}). In contrast to situations for options A and B, the expressions for the RR potentials are not very illuminating. 
\bigskip

As an alternative to the path discussed above, one can start from the result of option B and perform a duality along $h$ direction. In this case we find
\bea
&&(E-E\hat{N}^{-1}E)=\bar{g}_{ij}+\bar{b}_{ij}=\bar{\kappa}^a{}_i\bar{\kappa}^a{}_j+\epsilon_{ijk}\bar{b}_k,\quad \hat{N}\equiv( E+F+f_s)=(H+f_s)\nn
&& \bar{\zeta}^a=\bar{\kappa}^a{}_i\bar{z}^i\quad \bar{z}^i=\frac{(\bar{b}_i+\sqrt{2}v_i)}{\sqrt{\operatorname{det}\bar{g}}}
\eea
The matrix involved in dualization of the RR fields is
\bea
{\Omega_{nt_2}}^{-1}=(\bar{A}_0\Gamma^{\tilde{1}\tilde{2}\tilde{3}}+\bar{A}_{1a}\Gamma^{\tilde{a}})\Gamma_{11},\quad \bar{A}_0=\frac{1}{\sqrt{1+\bar{\zeta}^2}},\quad A_1=\frac{\bar{\zeta}^a}{\sqrt{1+\bar{\zeta}^2}},
\eea
and the dual frames are given by
\bea
&&\bar{e}=-\kappa X^{-T}(dv-\hat{N}^{-1}Eds)+\bar{\lambda} ds    \\
&&\bar{\lambda}=-\bar{\kappa}^{-1}\hat{N}^{-1}E,\quad X=[E-E\hat{N}^{-1}E+f_v],\quad {f_{s(v)}}_{ab}={s(v)}_kf^k_{ab}   \nonumber
\eea
For completeness we also give the explicit expressions for $\bar{\kappa}^a{}_i$ and other ingredients involved in the construction: 
\bea
\bar{\kappa}^a{}_i&=&\frac{1}{\gamma}\begin{pmatrix}
\alpha_1+\beta_1 s_1^2& \beta_1 s_1s_2&\beta_1 s_1s_3\\
\beta_1 s_1s_2&\alpha_1+\beta_1 s_2^2&\beta_1 s_2s_3\\
\beta_1 s_1s_3&\beta_1 s_2s_3&\alpha_1+\beta_1 s_3^2
\end{pmatrix} \nn
\gamma&=&\sqrt{Z}=\left[H(H^2+2{\bar r}^2)\right]^{\frac{1}{4}},\quad
\beta_1=-\frac{2 }{2\bar{r}^2}\sqrt{EH(FH+2{\bar r}^2)}+\sqrt{EFH^2+2\bar{r}^2}\nn
\alpha_1&=&\sqrt{EH(FH+2{\bar r}^2)},\quad \bar{r}^2=s_is^i\\
\sqrt{\operatorname{det}\bar{g}}&=&
\sqrt{\frac{(E^3 F)}{H}} \frac{(FH + 2\bar{r}^2)}{H^2 + 2\bar{r}^2},\quad b_k=\frac{E^2\sqrt{2}s_k}{H^2+2\bar{r}^2}\nn
\bar{z}^i&=&\sqrt{\frac{H}{E^3F}}\frac{(H^2+2\bar{r}^2)\sqrt{2}v^i+E^2\sqrt{2}s^i}{(FH+2\bar{r}^2)}\nonumber
\eea
The resulting RR fields reproduce (\ref{flux Gs A-1}).



\end{document}